\documentclass[
aps,
pra,,
nosuperscriptaddress,
twocolumn,
notitlepage
]{revtex4-2}
\usepackage{float}
\usepackage{graphicx}
\usepackage[svgnames,psnames]{xcolor}
\usepackage[colorlinks,
citecolor=DarkBlue,
linkcolor=DarkBlue,
urlcolor=DarkBlue,
unicode]{hyperref}

\usepackage{amsthm,amsmath,amssymb,amsbsy}
\usepackage{bbm}
\usepackage{physics}
\usepackage{siunitx}
\usepackage{mdframed}

\newcommand{\mc}[1]{\mathcal{#1}}
\newcommand{\ox}{\otimes}
\newcommand{\id}{\mathbbm{1}}
\newcommand{\spr}{\mathrm{spr}}

\newcommand{\kb}{k_\mathrm{B}}
\newcommand{\normord}[1]{:\mathrel{#1}:}

\newcommand{\inlineheading}[1]{\textit{#1}---\ignorespaces}

\newcommand{\coh}{\mc{C}}
\newcommand{\variance}{\mathrm{Var}}
\newcommand{\cov}{\mathrm{Cov}}
\newcommand{\nth}{\bar{n}_r}

\newcommand{\illum}{\mc{I}}

\newcommand{\gt}{g^{(2)}}
\newcommand{\ncov}{\widehat{\cov}}

\newtheorem{thm}{Theorem}
\newtheorem{lem}{Lemma}

\newtheorem{prop}{Proposition}

\begin{document} 

\title{Quantum phase sensing with states out of thermal equilibrium}

\author{Benjamin Yadin}

\affiliation{Naturwissenschaftlich-Technische Fakult\"at, Universit\"at Siegen, Walter-Flex-Straße 3, 57068 Siegen, Germany}
\email{benjamin.yadin@uni-siegen.de}
    
\begin{abstract}
    Interferometry can be viewed generally as the measurement of a relative phase between two subsystems.
    I consider the problem of interfering a quantum resource state with a thermal bath, drawing a precise connection between the athermality of the resource and the resulting phase sensitivity.
    This is done by finding the fundamental sensing precision limit under the minimal conditions of global unitarity and energy conservation.
    The results here apply both to general finite-dimensional systems and to linear quantum optics.
    The same techniques further upper-bound the speed at which a system and bath can jointly evolve under an energy-conserving interaction.
\end{abstract}

\date{\today}
\maketitle

\inlineheading{Introduction}
A system in a stationary state can produce motion by interacting with a thermal environment, as long as it is initially out of equilibrium.
This is familiar from the operation of heat engines using temperature gradients; it is intuitive that the larger the gradient, the faster the engine can run.
This illustrates the general principle in thermodynamics that an object out of equilibrium with a thermal background is a potential resource for doing something useful.
The resource viewpoint has been productive for the development of quantum thermodynamics in recent years, especially in a quantum information context~\cite{Janzing2000Thermodynamic,Horodecki2013Fundamental,Brandao2013Resource,Lostaglio2019Introductory}.

Quantum states that evolve quickly (under time or another parameter) are resources for metrology, the task of precisely sensing a physical parameter~\cite{Giovannetti2006Quantum,Paris2009Quantum,Toth2014Quantum,Degen2017Quantum}.
The limits imposed by thermodynamics on the precision and accurate timekeeping of quantum clocks is further an active area of exploration~\cite{Erker2017Autonomous,Guarnieri2019Thermodynamics,Pearson2021Measuring,Milburn2020Thermodynamics,Meier2023Fundamental,Prech2024Optimal,Meier2025Precision}.
An exemplary metrology setting is the use of optical interferometry to detect relative phase shifts between branches, notably for gravitational wave detection~\cite{Abadie2011Gravitational} among numerous other applications~\cite{Polino2020Photonic}.
Lasers are indispensible for interferometry due to their high coherence -- brightness in a small number of modes~\cite{Wiseman2016How} -- so are typically fed into one input port of an interferometer.
However, it is worth considering the second input port, which is generally assumed to be in the vacuum state.
This is justifiable at optical frequencies, as the thermal photon number is negligible: for example, at $\SI{300}{K}$, the mean number in a mode of wavelength $\SI{500}{n m}$ is $\bar{n} \sim  10^{-42}$.
In contrast, microwaves of wavelength $\SI{1}{c m}$ have $\bar{n} \sim 200$, and we may be forced to treat the second mode as thermal.
Then the brightness of the first input is insufficient to characterise the phase sensitivity of the interferometer.

In this paper, I show that athermality of the first input is the resource for phase sensing using interferometry in a thermal background.
Measuring performance by quantum Fisher information (QFI), this leads to a formal quantifier of athermality that can be thought of as extending the notion of optical coherence to a thermal setting.
For nonzero background temperature, this depends on statistical properties beyond mean photon number and can distinguish classical from nonclassical states.
This idea generalises to arbitrary finite-dimensional systems with any Hamiltonian, where I derive the optimal utility of a state for estimating a phase encoded through the system's Hamiltonian, given the ability to apply any energy-conserving unitary interaction with a thermal bath at a certain background temperature.
The phase sensitivity enters through system-environment correlations associated with superpositions of different system energies.
Hence the interaction effectively \emph{activates} the system's athermality into coherence.
This adds to the paradigm of activation of one quantum resource into another; previous examples involve mapping optical nonclassicality~\cite{Asboth2005Computable}, coherence~\cite{Streltsov2015Measuring}, discord~\cite{Piani2011All}, and identical particle entanglement~\cite{Killoran2014Extracting,Morris2020Entanglement} into bipartite entanglement.

The same methods solve the related problem of finding the maximal speed at which a system and thermal bath can interact.
Again, the bath is arbitrary (with fixed temperature), and the interaction is only constrained to conserve total energy and have bounded strength.
Speed here means square-root QFI, justified by its geometrical interpretation involving a state-space metric~\cite{Uhlmann1992Energy} and appearance in a Mandelstam--Tamm-type quantum speed limit~\cite{Mandelstam1945Uncertainty,Marvian2016Quantum}.
This result is hence a kind of speed limit set purely by the system's athermality.

Below, I first motivate the problem via interferometry while introducing the technical background, then present the results for general finite-dimensional systems, followed by a version for linear optical interferometers.

\inlineheading{Preliminaries}
The central concepts of this work are best illustrated in a standard Mach-Zehnder interferometer.
Two modes $A$ and $B$ of light are incident upon a balanced beam splitter: one mode described by state $\rho_A$, and the other in the vacuum state $\ket{0}_B$.
The encoding of a phase difference $\theta$ between the branches $C$ and $D$ after the beam splitter can be described by applying a unitary $e^{-i \theta N_C}$ to one branch, where the generator $N_C$ is the photon number observable in mode $C$.
A measurement on the resulting two-mode quantum state is made, typically by recombining the modes at another beam splitter and detecting intensity or photon number at the two output ports.
The difference in intensities can then be used to estimate the parameter $\theta$.
This measurement is insensitive to the phase of the input -- that is, invariant under a rotation of the state $\rho_A \to e^{-i \phi N_A} \rho_A e^{i \phi N_A}$.
More generally, any measurement that does not employ an external phase reference correlated with the input is phase-invariant.
Only the \emph{relative} phase between branches is measurable~\cite{Lang2013Optimal}.
We can regard the phase of $\rho_A$ as a random variable $\phi$ uniformly distributed over $[0,2\pi)$, which is broadcast into both branches by the beam splitter.
One branch acquires an additional shift $\theta$ that can be detected using the other branch as a phase reference.
The lack of an external phase reference imposes a superselection rule~\cite{Bartlett2005Dialogue,Bartlett2007Reference} that implies $\rho_A$ can always effectively be replaced by its average over all phases $\phi$ -- i.e., diagonal in the number basis, $\rho_A = \sum_{n=0}^\infty p_n \dyad{n}$.

The optimal precision in estimating $\theta$ for a given input state is found through the framework of quantum metrology~\cite{Giovannetti2006Quantum,Toth2014Quantum}.
The relevant problem is single-parameter estimation encoded via a fixed Hamiltonian $H$ -- we are required to estimate $\theta$ from measurements on $\rho(\theta) = e^{-i\theta H}\rho e^{i\theta H}$, where both $\rho$ and $H$ are known.
Suppose an observable $M$ is measured, giving rise to outcome $i$ following the probability distribution $p_M(i|\theta)$.
From data gathered over $m$ identical and independent measurement rounds, one calculates an estimate $\hat{\theta}$.
This should be unbiased, so its mean is the true value of the parameter: $\ev*{\hat{\theta}} = \theta$.
The miminum possible uncertainty, as captured by the mean-square difference from the correct value $(\Delta \theta)^2 = \ev*{(\hat{\theta}-\theta)^2}$, is determined by a key result in classical estimation theory, the Cram\'er-Rao bound: $(\Delta \theta)^2 \geq [m J_M(\theta)]^{-1}$~\cite{Cover2005ElementsCh11}.
Here, $J_M(\theta) = \ev*{[\partial_\theta \ln p_M(i|\theta)]^2}$ is the Fisher information, a measure of the sensitivity of the distribution $p_M(i|\theta)$ to variations in $\theta$.
The bound can be approached in limit of many measurement rounds.
Considering how to optimise the measurement $M$ to minimise the uncertainty leads to the quantum Cram\'er-Rao bound $(\Delta \theta)^2 \geq [m \mc{F}(\rho, H)]^{-1}$~\cite{Braunstein1994Statistical}, where $\mc{F}(\rho, H) = \max_M J_M(\theta)$ is known as the quantum Fisher information (QFI).
The central quantity considered in ths work, QFI is a measure of the sensitivity of the initial probe state $\rho$ to phase shifts applied by the generator $H$.
It also has a geometrical interpretation as the (squared) speed of evolution under the parameter $\theta$ according to a metric on state space~\cite{Uhlmann1992Energy,Marvian2016Quantum}.

Applying these tools to the Mach-Zehnder interferometer, we take the probe state $\sigma_{CD}$ just after the first beam splitter, and the generator $N_C$.
The QFI then turns out to be simply the mean photon number: $\mc{F}(\sigma_{CD}, N_C) = \ev{N_A} = \sum_{n=0}^\infty p_n n$~\cite{Lang2013Optimal,Jarzyna2015True,Takeoka2017Fundamental}.
In other words, the average energy determines phase sensitivity.
To clarify a possible point of confusion: for a beam of light, the maximum mean photon number in any frequency mode may be termed its ``coherence'', as this is connected to its optical temporal coherence~\cite{Baker2021Heisenberg}.
However, the diagonal state $\rho_A$ would be called incoherent in other literature~\cite{Baumgratz2014Quantifying,Marvian2016How} due to its lack of superposition in the energy eigenbasis.
The beam splitter acts to create superpositions of the form $\ket{n}_A \ket{0}_B \to \sum_k a_{n,k} \ket{n-k}_C \ket{k}_D$ which encode the relative phase.
Hence, we may instead think of the input as containing a kind of ``latent coherence'' -- a property which can be activated into coherent superpositions of different (branch-local) particle numbers in combination with a second vacuum mode and a beam splitter.

Now consider a natural generalisation of interferometry beyond optics.
The input systems $A,\,B$ have the same Hamiltonian $H_A=H_B = \sum_{i=0}^{d-1} \epsilon_i \dyad*{\epsilon_i}$.
System $A$ begins in some diagonal state $\rho_A = \sum_i p_i \dyad{\epsilon_i}$ and $B$ in its ground state $\ket{0}$.
Instead of a beam splitter, use an energy-conserving entangling unitary $U$ such that $U \ket{\epsilon_i} \ket{0} = (\ket{\epsilon_i}\ket{0} + \ket{0}\ket{\epsilon_i})/\sqrt{2}$.
A relative phase is then encoded by applying $e^{-i \theta H_A}$, mapping $\ket{\epsilon_i}\ket{0} + \ket{0}\ket{\epsilon_i} \to e^{-i \theta \epsilon_i} \ket{\epsilon_i}\ket{0} + \ket{0}\ket{\epsilon_i}$.
For each input $\ket{\epsilon_i}$, the resulting QFI is $\epsilon_i^2$; overall, we obtain $\mc{F}(U[\rho_A \ox \dyad{0}_B] U^\dagger, H_A) = \sum_i p_i \epsilon_i^2$.
This is analogous to the optical case, albeit with mean energy replaced by mean squared energy.

\begin{figure}[h]
    \includegraphics[width=.43\textwidth]{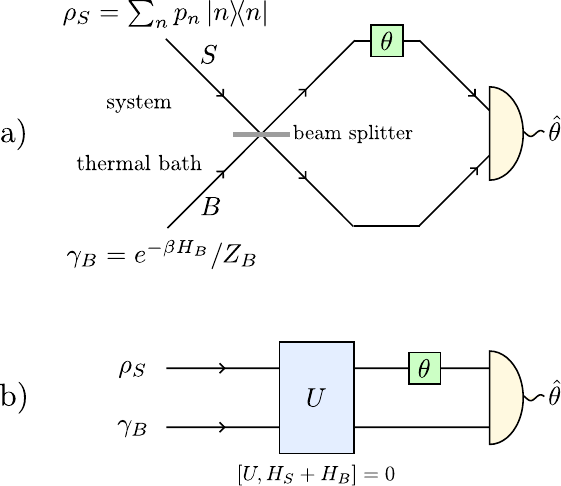}
    \caption{The main settings in this work: a) A phase-invariant state of light in mode $S$ is interfered with light from a thermal bath $B$. An encoded relative phase $\theta$ is estimated as $\hat{\theta}$ from a measurement. b) More generally, we consider an energy-conserving unitary interaction and a phase encoded by the system's Hamiltonian $H_S$.}
    \label{fig:setting}
\end{figure}

The main question addressed here is how to use a quantum system for phase sensing in combination with a \emph{thermal} background -- see Fig.~\ref{fig:setting}.

\inlineheading{Finite systems}
Consider a system $S$ of finite Hilbert space dimension with arbitrary Hamiltonian $H_S = \sum_{i=0}^{d-1} \epsilon_i \dyad{\epsilon_i}$, where the energies are $0 = \epsilon_0 \leq \epsilon_1 \leq \dots \leq \epsilon_{d-1}$.
This is combined with a thermal bath $B$ with Hamiltonian $H_B = \sum_{\eta \geq 0} \eta \sum_{\mu=0}^{D(\eta)-1} \dyad{\eta,\mu}$, where each energy $\eta$ has degeneracy $D(\eta)$.
The bath starts in the thermal state at a fixed temperature $T$: $\gamma_B = e^{-\beta H_B}/Z_B$, where $Z_B = \sum_\eta D(\eta) e^{-\beta \eta}$ is the partition function and $\beta = [\kb T]^{-1}$ is the inverse temperature.
The goal is to interact $S$ and $B$ such that the resulting state is sensitive to the phase encoding $e^{-i \theta H_S}$.
We allow any unitary $U$ that conserves total energy, so that $[U, H_S + H_B] = 0$.

This formulation follows the \emph{thermal operations} resource theory of quantum thermodynamics~\cite{Janzing2000Thermodynamic,Horodecki2013Fundamental,Brandao2013Resource}, whose aim is to account for all sources and sinks of energy and entropy by explicitly including all participating systems.
Unitarity is imposed so that the global dynamics are reversible, and energy must be conserved overall.
This describes scattering- or collision-type processes in which the system and bath are initially and finally decoupled~\cite{Strasberg2017Quantum,Ciccarello2022Quantum,Gaida2025Thermodynamically}.
Any thermal state at temperature $T$ is ``free'' -- may always be appended without associated cost -- while all resources out of equilibrium are included in the system $S$.

As above, we disallow external phase references, so $\rho_S = \sum_i p_i \dyad{\epsilon_i}$.
If $S$ is intially in thermal equilibrium with $B$, then the global state $\gamma_S \ox \gamma_B \propto e^{-\beta(H_S + H_B)}$ is unchanged under any energy-conserving $U$, so there is zero phase sensitivity.
For any $\rho_S \neq \gamma_S$, however, nonzero QFI may be possible.
We ask how to choose both a bath $B$ and interaction $U$ so as to maxmise the QFI of the global state -- formally, we define
\begin{align} \label{eqn:max_qfi_def}
    \mc{F}_T(\rho_S, H_S) := \sup_{H_B} \max_{\substack{U \text{,} \\ [U,H_S+H_B]=0}} \mc{F}\left( U\left[ \rho_S \ox \gamma_B \right] U^\dagger , H_S \right).
\end{align}
The task is thus to first find the best interaction for any given bath, and then to optimise over the bath's energy levels and degeneracies.
Note that optimisation over the bath is described by a supremum rather than a maximum as the bath may be arbitrarily large -- so the limit may not be attained exactly for any finite bath.
The first main result, the solution to this problem, will be stated and then explained via simple special cases.
A full derivation is given in Appendix~\ref{app:discrete_proof}.

\begin{figure}[h]
    \includegraphics[width=.45\textwidth]{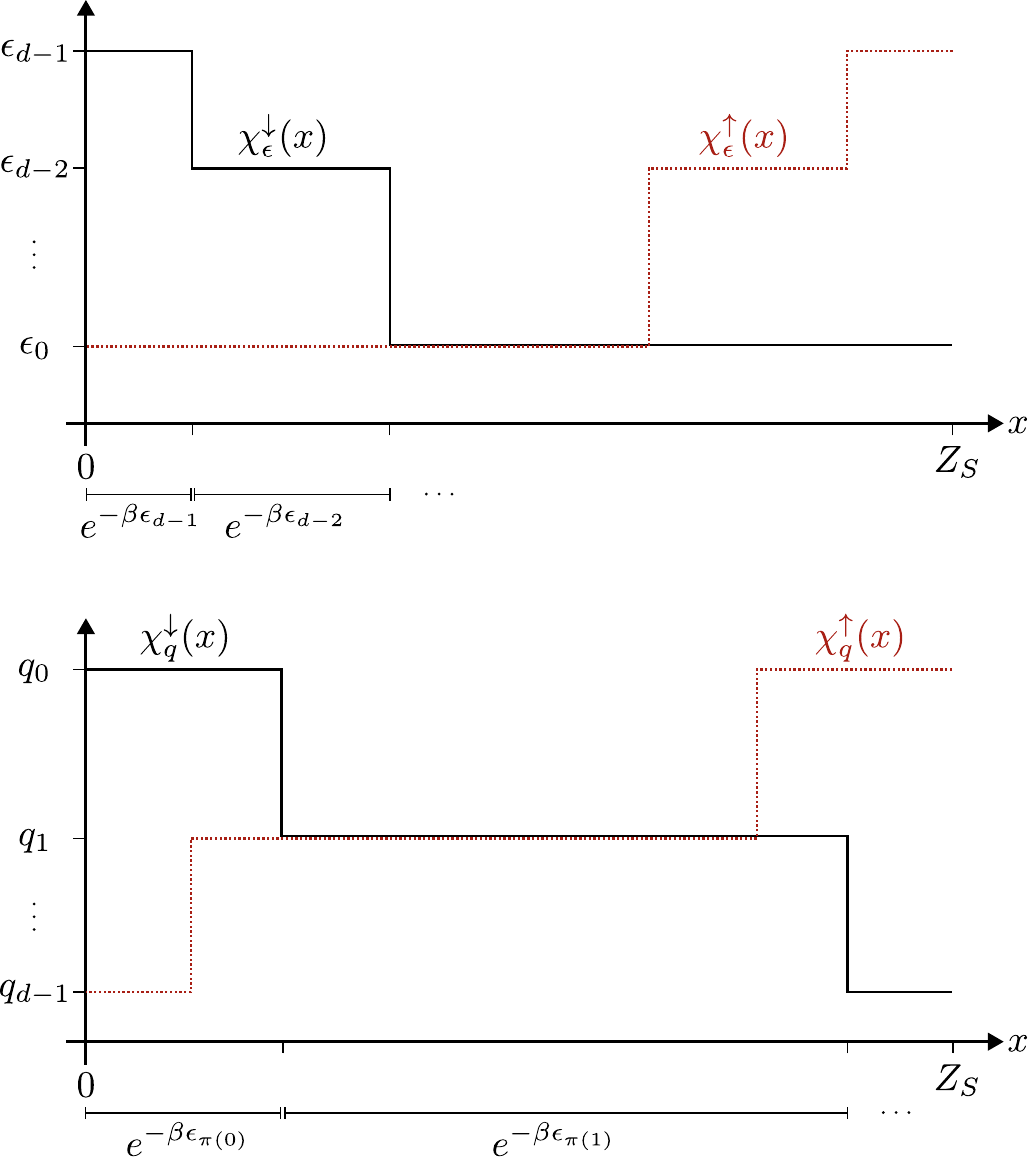}
    \caption{Illustration of the step functions used in Eq.~\eqref{eqn:max_qfi_result}.}
    \label{fig:steps}
\end{figure}

We need to construct two step functions of a variable $x$ that runs from zero to the partition function $Z_S$ -- see Fig.~\ref{fig:steps}.
The first function $\chi_\epsilon^\downarrow(x)$ takes the value of each system energy $\epsilon_i$, ordered decreasingly, in steps of size $e^{-\beta \epsilon_i}$.
So $\chi_\epsilon^\downarrow(x) = \epsilon_{d-1}$ (the largest energy) on the interval $[0, e^{-\beta \epsilon_{d-1}})$, then $\epsilon_{d-2}$ on $[e^{-\beta \epsilon_{d-1}}, e^{-\beta \epsilon_{d-1}} + e^{-\beta \epsilon_{d-2}} )$, and so on.
The reflection of this function in the $x$-axis about its midpoint $Z_S/2$ is denoted $\chi_\epsilon^\uparrow(x) = \chi_\epsilon^\downarrow(Z_S-x)$.
Secondly, we weight the probabilities by inverse Boltzmann factors $p_i e^{\beta \epsilon_i}$ and find a permutation $\pi$ of indices such that $q_i := p_{\pi(i)} e^{\beta \epsilon_{\pi(i)}}$ are ordered decreasingly, i.e., $q_0 \geq q_1 \geq \dots \geq q_{d-1}$ (known as $\beta$-ordering~\cite{Horodecki2013Fundamental}).
The function $\chi_q^\downarrow(x)$ is then constructed similarly by taking the values $q_0,\, q_1,\, \dots$ in order, in steps of size $e^{-\beta \epsilon_{\pi(i)}}$.
The optimal QFI is then expressed as
\begin{align} \label{eqn:max_qfi_result}
    \mc{F}_T(\rho_S,H_S) = \frac{1}{2}\int_0^{Z_S} \dd x \; f(\chi_q^\downarrow[x], \chi_q^\uparrow[x])(\chi_\epsilon^\downarrow[x] - \chi_\epsilon^\uparrow[x])^2,
\end{align}
where we term the function $f(a,b) := (a-b)^2 /(a+b)$ the \emph{Fisher difference}.
This quantity is symmetric, vanishes if and only if $a=b$, and satisfies $f(ca, cb) = c f(a,b)$ and $f(a,0) = a$.
(Note that the integral is symmetric about the midpoint $x = Z_S/2$.)

The integrand in Eq.~\eqref{eqn:max_qfi_result} involves the squared difference of pairs of energies and the Fisher difference of pairs of Boltzmann-weighted probabilities.
$\mc{F}_T$ vanishes precisely when $p_i e^{\beta \epsilon_i} = p_j e^{\beta \epsilon_j}$ for every pair of energy levels -- equivalent to $\rho_S = \gamma_S$.
Moreover, $\mc{F}_T$ quantitatively measures the athermality of $\rho_S$, since it is by construction monotone decreasing under thermal operations.
In other words, letting the system partially thermalise by first interacting with some other bath $B'$, $\rho_S \to \rho'_S = \tr_{B'} [ U' (\rho_S \ox \gamma_{B'}) {U'}^\dagger ]$, always results in $\mc{F}_T(\rho'_S, H_S) \leq \mc{F}_T(\rho_S, H_S)$.

Since Eq.~\eqref{eqn:max_qfi_result} is an integral of a step function, it can always in principle be written as a sum; however, the precise form depends on the ordering of energies and probabilities.
It has a notable simplification when $S$ is a qubit, when $\rho_S = p_0 \dyad{0} + p_1 \dyad{\epsilon}$ and $H_S = \epsilon \dyad{\epsilon}$.
On the left half of the range, the factor $(\chi_\epsilon^\downarrow[x] - \chi_\epsilon^\uparrow[x])^2$ is nonzero only on $x \in [0, e^{-\beta \epsilon})$, and where the other factor is nonzero, it always equals $f(p_0, p_1 e^{\beta \epsilon})$ independent of the $\beta$-ordering.
Therefore, the qubit case is
\begin{align} \label{eqn:qfi_qubit}
    \mc{F}_T^\text{(qubit)} = e^{-\beta \epsilon} f(p_0, p_1 e^{\beta \epsilon}) \epsilon^2 = f(p_0 e^{-\beta \epsilon}, p_1) \epsilon^2.
\end{align}
The low- and high-temperature limits are $\mc{F}_0^\text{(qubit)} = p_1 \epsilon^2$ and $\mc{F}_\infty^\text{(qubit)} = (2p_1 - 1)^2 \epsilon^2$, scaling respectively with excitation probability and purity.

For arbitrary system dimension, these temperature limits can also be found:
\begin{align} \label{eqn:qfi_temp_limits}
    \mc{F}_0 = \sum_{i=0}^{d-1} p_i \epsilon_i^2 , \quad \mc{F}_\infty = \sum_{i=0}^{(d-1)/2} f( p^\downarrow_i, p^\uparrow_i) (\epsilon^\downarrow_i - \epsilon^\uparrow_i)^2,
\end{align}
where the arrow notation indicates increasing or decreasing order: $p^\uparrow_0 \leq p^\uparrow_1 \leq \dots$, $p^\downarrow_0 \geq p^\downarrow_1 \geq \dots$, and so on.
The low-temperature limit recovers the example discussed in the previous section where the bath is its ground state.
The high-temperature limit gives a quantity that captures purity analogously to the negative of entropy: it vanishes when $\rho_S$ is maximally mixed and is maximised for any pure state.
Moreover, this is the same maximal QFI that could be obtained by instead rotating the system alone under \emph{any} (not energy-conserving) unitary: $\mc{F}_\infty(\rho_S,H_S) = \max_{\text{all } U} \mc{F}(U \rho_S U^\dagger, H_S)$~\cite{Fiderer2019Maximal}.
These observations exemplify the principle that energy and entropy are respectively the relevant properties at low and high temperature.
At any finite temperature, the optimal state is always the maximally excited state $\ket{\epsilon_{d-1}}$, giving $\mc{F}_T = \epsilon_{d-1}^2$.

The main idea behind the derivation of Eq.~\eqref{eqn:max_qfi_result} is that any energy-conserving $U$ takes a block-diagonal structure $U = \bigoplus_E U_E$ with respect to subspaces of fixed total energy $E$, and each $U_E$ can be varied arbitrarily within that subspace.
Thus, we can employ a maximum QFI result from Ref.~\cite{Fiderer2019Maximal}, the remaining problem being to vary the bath degeneracies, which determine the number of repetitions of system energies and probabilites in each subspace.
The optimal $D(E)$ scales exponentially in energy, $D(E + \epsilon) = D(E) e^{\beta \epsilon}$.
While impossible for any finite bath, this form can be approximated arbitrarily well for the most probable energies.
Such a bath is also sufficient for constructing the most general transformations of diagonal states under thermal operations~\cite{Horodecki2013Fundamental}.
In the qubit case, it suffices to take $B$ as a single bosonic mode resonant with the energy gap of $S$.
Note that the optimal interaction $U$ can always be explicitly described.

\inlineheading{Linear quantum optics}
Returning to the optical setting, we now consider a constrained version of the above question, where $S$ is a single bosonic mode of frequency $\omega$, and $B$ is composed of arbitrarily many modes of the same frequency.
The interaction $U \in \mathrm{LO}$ is permitted to be any (passive) linear optical unitary.
If $S$ has annihiliation operator $a_0$ and $B$ has $a_i$ for $i = 1,\dots,k$, then $U$ acts as $U^\dagger a_i U = \sum_{j=0}^k u_{ij} a_j$, where $u$ is a unitary matrix.
This setup allows for any network of beam splitters and phase shifters, and fully describes the restriction of the set of thermal operations to its Gaussian subset (considering only degenerate modes)~\cite{Serafini2020Gaussian,Narasimhachar2021Thermodynamic}.

In contrast to the finite-dimensional case, we find that a single bath mode with a balanced beam splitter is always sufficient to achieve the maximal QFI, given by
\begin{align} \label{eqn:max_qfi_optics}
    \sup_k \max_{U \in \mathrm{LO}} \mc{F}(U[\rho_S \ox \gamma^{\ox k}_B]U^\dagger, H_S) & = (\hbar \omega)^2 (\nth + 1) \coh_r(\rho_S),
\end{align}
where we have factored out the mode frequency and a dependence on the mean thermal photon number $\nth = r/(1-r)$, $r = e^{-\beta \hbar \omega}$.
The dependence on $\rho_S = \sum_n p_n \dyad{n}$ is termed here the \emph{latent coherence},
\begin{align}
    \coh_r(\rho_S) := \sum_{n=1}^\infty f(p_n, r p_{n-1}) n.
\end{align}
$\coh_r$ is again by construction a monotone under Gaussian thermal operations and vanishes exactly when $\rho_S = \gamma_S$.
The low-temperature limit recovers $\coh_0(\rho) = \ev{N_S}$ as expected.
At high temperature, we have $\coh_1(\rho) = \frac{1}{2}\mc{F}(\rho_S, x)$, where $x = (a + a^\dagger)/\sqrt{2}$ is a phase space quadrature.
This means that high-temperature performance is equivalent to sensitivity under linear displacements in phase space, known to be an indicator~\cite{Rivas2010Precision,Kwon2019Nonclassicality,Yadin2018Operational} of Glauber-Sudarshan ($P$-function) nonclassicality~\cite{Glauber1963Quantum,Sudarshan1963Equivalence}.
In fact, $\coh_r$ provides a nonclassicality witness for every $r>0$: if $\rho_\text{cl}$ is classical, then
\begin{align}
    \coh_r(\rho_\text{cl}) \leq (1-r) \ev{N}  + r.
\end{align}
The general upper bound for all states is instead
\begin{align} \label{eqn:max_coh_fix_n}
    \coh_r(\rho) \leq (1+r) \ev{N} + r
\end{align}
and is attained whenever subtracting a photon leaves an orthogonal state, i.e., $\tr[\rho a \rho a^\dagger] = 0$.
For example, $\coh_1(\rho_\text{cl}) \leq 1$ while the maximum of $\coh_1$ in Eq.~\eqref{eqn:max_coh_fix_n} is $2\ev{N}$.
Only the zero-temperature case $\coh_0$ fails to witness nonclassicality.
Other properties are listed in Appendix~\ref{app:max_qfi_optical}.

We evaluate $\coh_r$ for some common quantum optical states (see Appendix~\ref{app:optics_examples}).
A Fock state $\ket{n}$ achieves the maximal value $(1+r)n+r$, a thermal state $\gamma'$ at a different temperature $T'$ has $(r'-r)^2/(r'+r)(1-r')$, and a state $\rho_\lambda$ with Poisson statistics $p_n = e^{-\lambda} \lambda^n/n!$ with mean $\lambda$ has an integral representation
$\coh_r(\rho_\lambda) = r(\lambda+1) -3\lambda + 4\frac{\lambda}{r} - 4 \frac{\lambda^2}{r^2} \int_0^1 \dd u \; u^{\lambda/r-1} e^{\lambda(u-1)}$.
Furthermore, some connections with standard optical coherence functions are outlined in Appendix~\ref{app:standard_coherence}.

The utility for phase sensing in a thermal background exists in a trade-off with performance in a quantum illumination protocol~\cite{Lloyd2008Enhanced}, which can be cast in metrological terms as sensing a weakly reflected signal on top of a noisy background~\cite{Sanz2017Quantum}.
This task can be modelled as mixing a signal with a thermal background at a beam splitter with reflectivity $R \ll 1$; one needs to estimate the parameter $R$.
With a single signal mode, the sensitivity vanishes as $R \to 0$, but keeping it entangled with an ``idler'' $I$ gives a finite QFI~\cite{Sanz2017Quantum} that can be written as $\mc{I}_r(\rho) / \nth$, where
$\mc{I}_r(\rho) := 4 \sum_{n=1}^\infty \frac{r p_n p_{n-1}}{p_n + rp_{n-1}} n$.
It is easy to show that $\coh_r(\rho) + \mc{I}_r(\rho) = (1+r) \ev{N} + r$, so performance in one estimation task is traded for the other, given a certain mean photon number.
Details and a physical interpretation are given in Appendix~\ref{app:illumination}.

\inlineheading{Interaction speed}
A small variation of our methods allows us to answer a different question: How quickly can a system and thermal bath interact?
Instead of applying a fixed unitary, consider an interaction Hamiltonian $H_I$ and time evolution $U_t = e^{-itH_{SB}/\hbar}$ under $H_{SB} = H_S + H_B + H_I$.
As before, we require the interaction to conserve total energy of the parts, so $[H_I, H_S + H_B] = 0$.
The QFI under the generator $H_{SB}$ reduces to that under just $H_I$ due to the diagonality of the initial state; hence, we ask for the quantity
\begin{align} \label{eqn:max_qfi_int_def}
    \mc{F}^\text{int}_T(\rho_S, H_S) := \sup_{H_B} \max_{H_I} \, \mc{F}\left( \rho_S \ox \gamma_B , H_I \right).
\end{align}
In addition to energy conservation, we must constrain the size of $H_I$ for this to be well-defined.
A convenient choice is to fix the \emph{spread} of each energy block $\spr(H_{I,E}) \leq 1$, where $H_I = \bigoplus_E H_{I,E}$.
The spread of a Hermitian operator $A$ is the difference between its largest and smallest eigenvalues: $\spr(A) = \lambda_{\max}(A) - \lambda_{\min}(A)$~\cite{Mirsky1956Spread}.
The result is a simpler version of Eq.~\eqref{eqn:max_qfi_result} (see Appendix~\ref{app:speed_discrete})
\begin{align} \label{eqn:max_qfi_int_result}
    \mc{F}^\text{int}_T(\rho_S,H_S) & = \frac{1}{2} \int_0^{Z_S} \dd x \; f\left(\chi_q^\downarrow[x], \chi_q^\uparrow[x]\right) .
\end{align}

The same process can be followed for the optical case (Appendix~\ref{app:speed_optical}), where we now express the interaction as a general bilinear Hamiltonian $H_I = \vb{a}^\dagger h \vb{a} = \sum_{i,j=0}^k h_{ij} a_i^\dagger a_j$ and set $\spr(h) \leq 1$.
This in fact turns out to be identical to the phase sensing result Eq.~\eqref{eqn:max_qfi_optics}, obtained for a single bath mode and a beam splitter Hamiltonian $H_I = (a_0^\dagger a_1 + a_1^\dagger a_0)/2$.

These results have implications for quantum speed limits in open systems~\cite{Deffner2017Quantum}, where one is interested in the evolution of the system alone, without access to the bath.
The system's state after an interaction time $t$ is $\rho_S(t) = \tr_B [U_t (\rho_S \ox \gamma_B) U_t^\dagger]$.
Due to the information-processing property, the QFI of the family of states $\rho_S(t)$ under the parameter $t$ is never larger than the QFI of the total state, so $\mc{F}[\rho_S(t)] \leq \mc{F}^\text{int}_T(\rho_S, H_S)$.
Importantly, this holds at every $t \geq 0$, so does not rely on Markovianity.
We only assume that the interaction is energy-conserving and that $S$ and $B$ are initially uncorrelated.
Whether the inequality is tight enough to be useful for practical examples is an open question.
Speed limits for this kind of setting have been derived in, for instance, Refs.~\cite{Deffner2013Quantum,Ilin2024Quantum}; however, those results require some knowledge of the interaction or the master equation, and do not make a connection to the athermality of the system.

\inlineheading{Conclusion}
In summary, we have seen that the ability of a quantum system to enocde a phase through interaction with a thermal background is a function of its athermality.
Note that we assumed an intially phase-invariant state, so this property is different from athermality due to coherence in the energy eigenbasis, which can be also measured with QFI~\cite{Janzing2003Quasi,Yadin2016General,Marvian2022Operational} or with relative entropy~\cite{Lostaglio2015Description}.
Such coherence is the topic of recent works on quantum thermodynamics~\cite{Gour2022Role,Tajima2025Gibbs,Shiraishi2024Quantum}, which implicitly include an ideal external phase reference.
Remarkably, while the athermality we encounter here is arguably classical -- being a function of the energy probability distribution -- it is intrinsically defined through a quantum protocol in which coherence is established via a relative phase.
Furthermore, this approach provides a certain unification of quantum optical coherence with discrete-system coherence.

There are many avenues for developing these ideas -- for example, considering the family of generalised QFI quantities~\cite{Petz2002Covariance}, one of which (the ``right logarithmic derivative'') places limits on coherence distillation~\cite{Marvian2020Coherence}.
Other extensions include multiple copies of resource states and multiple modes in the optical case, including different frequencies.
What role is played by correlations between modes, for instance, as found between narrowly filtered frequencies in laser light~\cite{Wiseman2016How}?
Moreover, these results may have implications for the thermodynamical limitations on clock precision~\cite{Erker2017Autonomous}, where the ultimate quantum bounds are still unknown~\cite{Meier2025Precision}.\\

\inlineheading{Acknowledgments}
I thank Gerardo Adesso and Stefan Nimmrichter for valuable discussions.
This project has received funding from the European Union's Horizon 2020 research and innovation programme under the Marie Sk\l odowska-Curie grant agreement No.~945422.
This work has been supported by the Deutsche Forschungsgemeinschaft (DFG, German Research Foundation, project numbers 447948357 and 440958198), the Sino-German Center for Research Promotion (Project M-0294), and the German Ministry of Education and Research (Project QuKuK, BMBF Grant No.~16KIS1618K).

\bibliography{main}

\appendix

\onecolumngrid

\section{QFI and Fisher difference properties} \label{app:fisher_diff}

For a single parameter $\theta$, the QFI of a family of states $\rho(\theta)$ is a function of both the state at $\theta$ and its derivative $\partial_\theta \rho(\theta)$.
One expression, in terms of the eigenvalues $p_i(\theta)$ and eigenstates $\ket{\psi_i(\theta)}$ of $\rho(\theta)$, is \cite{Paris2009Quantum}
\begin{align}
    \mc{F}[\rho(\theta)] = 2 \sum_{i,j} \frac{\abs{\mel{\psi_i(\theta)}{\partial_\theta \rho(\theta)}{\psi_j(\theta)}}^2}{p_i(\theta)+p_j(\theta)},
\end{align}
where terms in the sum with $p_i(\theta)=p_j(\theta)=0$ are excluded.
In this work, we mainly focus on the unitary case with a constant Hamiltonian generator $H$, where $\rho(\theta) = e^{-i\theta H} \rho e^{i\theta H}$.
Then, the QFI is the same for any $\theta$ and is written as a function of $\rho$ and $H$:
\begin{align} \label{eqn:qfi_expression}
    \mc{F}(\rho,H) & = 2 \sum_{i,j} \frac{(p_i-p_j)^2}{p_i+p_j} \abs{\mel{\psi_i}{H}{\psi_j}}^2  \\
        & = 2 \sum_{i,j} f(p_i,p_j) \abs{\mel{\psi_i}{H}{\psi_j}}^2 ,
\end{align}
again excluding $p_i=p_j=0$ terms.\\

Some important properties of QFI include~\cite{Petz2002Covariance}:
\begin{enumerate}
    \item Scaling with constants: For any constant $c\geq 0$, $\mc{F}[c \rho(\theta)] = c\mc{F}[\rho(\theta)]$.
        Scaling with the Hamiltonian is instead quadratic: for any $c \in \mathbb{R}$, $\mc{F}(\rho,cH) = c^2 \mc{F}(\rho,H)$.
    \item Convexity: For any pair of parameterised states $\rho(\theta),\, \sigma(\theta)$ and probability $p$, $\mc{F}[p\rho(\theta)+(1-p)\sigma(\theta)] \leq p\mc{F}[\rho(\theta)] + (1-p)\mc{F}[\sigma(\theta)]$.
        In the unitary case, this implies $\mc{F}(p\rho+[1-p]\sigma,H) \leq p\mc{F}(\rho,H) + (1-p)\mc{F}(\sigma,H)$.
    \item Monotonicity: For any quantum channel (trace-preserving completely positive map) $\Lambda$, $\mc{F}[\Lambda(\rho\{\theta\})] \leq \mc{F}[\rho(\theta)]$.
        A special statement for the unitary case holds for \emph{translationally invariant} channels~\cite{Marvian2016How}, i.e., those commuting with the unitary evolution such that $\Lambda (e^{-i\theta H} \rho e^{i\theta H}) = e^{-i\theta H} \Lambda(\rho) e^{i\theta H} \; \forall \rho$.
        For such channels, we have $\mc{F}(\Lambda[\rho],H) \leq \mc{F}(\rho,H)$.
    \item Additivity under a block-diagonal structure: Suppose there exists a basis in which both $\rho$ and $H$ take the same block-diagonal form, $\rho = \bigoplus_k \rho_k$, $H = \bigoplus_k H_k$.
        Then $\mc{F}(\rho,H) = \sum_k \mc{F}(\rho_k, H_k)$.
\end{enumerate}

We will furthermore make use of certain properties of the Fisher difference (some of which are closely related to the above):
\begin{prop} \label{prop:fisher_diff}
    \begin{mdframed}
        The Fisher difference $f(x,y) := (x-y)^2/(x+y)$ satisfies the following:
        \begin{enumerate}
            \item $f(x,y) = f(y,x)$;
            \item $f(cx, cy) = c f(x,y)$, therefore we can also define $f(0,0)=0$;
            \item $f(x,0) = x$;
            \item $f(x,y) = 0$ if and only if $x = y$;
            \item $\partial_x f(x, y) > 0$ if $x > y \geq 0$;
            \item $f(x,y)$ is jointly convex, and therefore satisfies
                \begin{align*}
                     f(px + [1-p] x', py + [1-p] y') \leq pf(x,y) + (1-p)f(x',y') \quad \forall \, p \in [0,1].
                \end{align*}
        \end{enumerate}
    \end{mdframed}
\end{prop}
Properties (1-4) are easy to see.
(5) follows immediately from the derivative
\begin{align}
    \partial_x f(x,y) & = \frac{(x-y)(x+3y)}{(x+y)^2}.
\end{align}
Finally, for (6) we compute the second derivatives
\begin{align}
    \partial_x^2 f(x,y) = \frac{8 y^2}{(x+y)^3}, \quad \partial_y^2 f(x,y) = \frac{8 x^2}{(x+y)^3}, \quad \partial_x \partial_y f(x,y) = \frac{-8 xy}{(x+y)^3},
\end{align}
giving a nonnegative Hessian matrix $[\partial_i \partial_j f]_{i,j=x,y} \geq 0$.

\section{Maximal QFI in finite dimensions} \label{app:discrete_proof}

\subsection{Main result} \label{app:discrete_main}

Here, we prove the main result on the maximal QFI for phase sensing in combination with a thermal bath:
\begin{thm} \label{thm:max_qfi_discrete}
    \begin{mdframed}
        For any $d$-dimensional system with Hamiltonian $H_S = \sum_{i=0}^{d-1} \dyad{\epsilon_i}$ in the state $\rho_S = \sum_{i=0}^{d-1} p_i \dyad{\epsilon_i}$, the maximal phase-sensing QFI relative to any thermal bath at temperature $T$ is
        \begin{align}
            \mc{F}_T(\rho_S, H_S) = \int_0^{Z_S/2} \dd x \; f(\chi_q^\downarrow[x], \chi_q^\uparrow[x]) (\chi_\epsilon^\downarrow[x]-\chi_\epsilon^\uparrow[x])^2,
        \end{align}
        where $\chi_\epsilon^{\downarrow,\uparrow},\, \chi_q^{\downarrow,\uparrow}$ are as defined in the main text.
    \end{mdframed}
\end{thm}

We will need the following result from Ref.~\cite{Fiderer2019Maximal} on the maximal QFI for a given spectrum:
\begin{lem} \label{lem:max_qfi_unitary}
    Given a state $\rho = \sum_i p_i \dyad{\psi_i} $ with eigenvalues $p_0 \geq p_1 \geq \dots \geq p_{d-1}$ and a Hamiltonian $H = \sum_i \epsilon_i \dyad{\epsilon_i}$ with eigenvalues $\epsilon_0 \leq \epsilon_1 \leq \dots \leq \epsilon_{d-1}$, the maximal QFI under any unitary $U$ is
    \begin{align}
        \max_U \, \mc{F}(U \rho U^\dagger, H) & = \frac{1}{2} \sum_{i=0}^{d-1} f(p_i, p_{d-1-i}) (\epsilon_{d-1-i} - \epsilon_i)^2 \\
            & = \frac{1}{2} \sum_{i=0}^{d-1} f(p^\downarrow_i, p^\uparrow_i) (\epsilon^\downarrow_i - \epsilon^\uparrow_i)^2.
    \end{align}
    Note that the sum is symmetric about the midpoint, so one can instead sum from $i=0$ to $(d-1)/2$ (rounded down to the nearest integer) with a factor of two.
    An optimal unitary achieving this maximum rotates the eigenstates $\ket{\psi_i}$ of $\rho$ according to
    \begin{align}
        U \ket{\psi_i} =
        \begin{cases}
            \frac{1}{\sqrt{2}} (\ket{\epsilon_i} + \ket{\epsilon_{d-1-i}}), & i < (d-1)/2 \\
            \ket{\epsilon_i}, & i = (d-1)/2 \\
            \frac{1}{\sqrt{2}} (\ket{\epsilon_i} - \ket{\epsilon_{d-1-i}}), & i > (d-1)/2 
        \end{cases}.
    \end{align}
\end{lem}
Intuitively, this result says that the best case is to associate the highest probabilities with superpositions of the most different energies, balanced with the lowest probabilities mapped to superpositions with the opposite sign, such that the mixture maintains maximal coherence.
The Fisher difference $f(x,y)$ is maximised when $x$ and $y$ are as different as possible, which is why $p_0$ is paired with $p_{d-1}$, and so on.\\

To begin, we express the system and bath Hamiltonians as
\begin{align}
    H_S & = \sum_{i=0}^{d-1} \epsilon_i \dyad{\epsilon_i}, \\
    H_B & = \sum_{\eta \geq 0} \eta \sum_{\mu = 0}^{D(\eta) - 1} \dyad{\eta, \mu}.
\end{align}
Note that, for convenience, we label the system's energy eigenstates by their index but leave the number of bath energies $\eta$ open; we set $D(\eta) = 0$ if $\eta$ does not appear in the sum.
The initial product state and the interaction unitary are decomposed into blocks of total energy $E$:
\begin{align}
    \rho_S \ox \gamma_B & = \bigoplus_E P_E\,  \sigma_{SB|E}, \\
    U & = \bigoplus_E U_E,
\end{align}
where $P_E$ is the total probability of having energy $E$.
The above form of $U$ is the most general one for $U$ given $[U, H_S + H_B] = 0$ -- each $U_E$ can be any unitary on its block of dimension $\Delta_E$.
Therefore, we can apply Lemma~\ref{lem:max_qfi_unitary} independently to each block.

Given $\rho_S = \sum_{i=0}^{d-1} p_i \dyad{\epsilon_i}$, we have
\begin{align} \label{eqn:state_e_block}
    P_E \, \sigma_{SB|E} = \sum_{i=0}^{d-1} \sum_{\mu=0}^{D(E-\epsilon_i)-1} p_i \frac{e^{-\beta(E-\epsilon_i)}}{Z_B} \dyad{\epsilon_i}_S \ox \dyad{E-\epsilon_i, \mu}_B.
\end{align}
Up to normalisation, this conditional state has eigenvalues $p_i e^{\beta \epsilon_i}$, each repeated $D(E-\epsilon_i)$ times.
We therefore introduce the vector $\vb{q}_{|E}$, defined by placing these values in decreasing order:
\begin{align}
    \vb{q}_{|E} := (p_i e^{\beta \epsilon_i} \; [D(E-\epsilon_i) \text{ times}] \, )^\downarrow_i.
\end{align}
The eigenvalues of $H_S \ox \id_B$, projected onto the energy $E$ block, are
\begin{align}
    \vb*{\epsilon}_{|E} := (\epsilon_i \; [D(E-\epsilon_i) \text{ times}] \, )_i,
\end{align}
still in increasing order.

It follows from Lemma~\ref{lem:max_qfi_unitary} that
\begin{align} \label{eqn:qfi_e_block}
    \max_{U_E} \, \mc{F}(P_E \, U_E \sigma_{SB|E} U_E^\dagger, H_S) = \frac{e^{-\beta E}}{Z_B} \sum_{k=0}^{(\Delta_E-1)/2} f(q_{k|E}, q_{\Delta_E-1-k|E}) (\epsilon_{\Delta_E-1-k|E} - \epsilon_{k|E})^2.
\end{align}

Due to the block-diagonal structure implying $\mc{F}(\bigoplus_E P_E \sigma_{SB|E}, H_S) = \sum_E \mc{F}(P_E \sigma_{SB|E}, H_S)$, the total QFI is
\begin{align} \label{eqn:qfi_sum_blocks}
    \max_U \, \mc{F}(U \sigma_{SB} U^\dagger, H_S) & = \sum_E \max_{U_E} \, \mc{F}(P_E \, U_E \sigma_{SB|E} U_E^\dagger, H_S) \nonumber \\
        & = \sum_E \frac{e^{-\beta E}}{Z_B} \sum_{k=0}^{(\Delta_E-1)/2} f(q_{k|E}, q_{\Delta_E-1-k|E}) (\epsilon_{\Delta_E-1-k|E} - \epsilon_{k|E})^2.
\end{align}

The task now becomes to optimise the degeneracies $D(\eta)$ so as to maximise the expression in Eq.~\eqref{eqn:qfi_sum_blocks}.
How to do this is not at all obvious from inspecting the expression, so our logic proceeds in a different way:
\begin{itemize}
    \item Suppose we have some bath $B$ with Hamiltonian $H_B$. Then we can always append any additional bath $B'$, such that the new combined bath $\tilde{B} = BB'$, with Hamiltonian $H_B + H_{B'}$, has a maximal QFI at least as large. This is because $\max_{V_{SBB'}} \mc{F}(V [\rho_S \ox \gamma_B \ox \gamma_{B'}] V^\dagger, H_S)$ involves varying over a superset of the unitaries in $\max_{U_{SB}} \mc{F}(U [\rho_S \ox \gamma_B] U^\dagger, H_S)$. To see this, note that if $[U_{SB},H_S+H_B] = 0$, then $[U_{SB}\ox \id_{B'}, H_S+H_B+H_{B'}]=0$. Also, tracing out $B'$ cannot increase the QFI.
    \item We choose a specific $B'$ such that the degeneracies $\tilde{D}(E)$ of the combined $\tilde{B}$ scale exponentially, $\tilde{D}(E + \epsilon) \approx \tilde{D}(E) e^{\beta \epsilon}$, to arbitrary accuracy on the typical energies $E$ that are almost certainly occupied. The proof that this is possible is given in Lemma~\ref{lem:large_bath} in Appendix~\ref{app:large_bath}.
    \item It follows that an idealised (infinite) bath that achieves the maximum QFI can be assumed to have $D(E + \epsilon) = D(E) e^{\beta \epsilon}$ for all $E$.
\end{itemize}

So we now evaluate Eq.~\eqref{eqn:qfi_sum_blocks} under the assumption $D(E + \epsilon) = D(E) e^{\beta \epsilon}$.
Each distinct term corresponding to system energy $\epsilon_i$ in $\vb{q}_{|E}$ and $\vb*{\epsilon}_{|E}$ is repeated $D(E)e^{-\beta \epsilon_i}$ times, where $D(E)$ is a sufficiently large number that $D(E) e^{-\beta \epsilon_i} \gg 1 \; \forall \, i$.
Thus, $\epsilon_{k|E}$, as a function of $k$, looks like a step function taking the increasing values $\epsilon_i$ in steps of size $D(E)e^{-\beta \epsilon_i}$.
Rescaling the $x$-axis by $k \to x = k/D(E)$ thus recovers the step function defined in the main text, such that $\epsilon_{\Delta_E-1-k|E}$ is replaced by $\chi_\epsilon^\downarrow(x)$ and $\epsilon_{k|E}$ by $\chi_\epsilon^\uparrow(x)$.
The same idea holds for $q_{k|E}$ and $\chi_q^\downarrow(x)$.
The sum over $k$ approaches an integral over $x$ in which $\dd x$ is compensated by the scaling factor $D(E)$.
Therefore,
\begin{align}
    \mc{F}_T(\rho_S, H_S) & = \sum_E \frac{e^{-\beta E}}{Z_B} D(E) \int_0^{Z_S/2} \dd x \; f(\chi_q^\downarrow[x], \chi_q^\uparrow[x]) (\chi_\epsilon^\downarrow[x]-\chi_\epsilon^\uparrow[x])^2 \nonumber \\
        & = \int_0^{Z_S/2} \dd x \; f(\chi_q^\downarrow[x], \chi_q^\uparrow[x]) (\chi_\epsilon^\downarrow[x]-\chi_\epsilon^\uparrow[x])^2,
\end{align}
since $Z_B = \sum_E e^{-\beta E} D(E)$. \\

The unitary $U_E$ that achieves the maximum in Eq.~\eqref{eqn:qfi_e_block} can also be found from Lemma~\ref{lem:max_qfi_unitary}.
The eigenstates of $\sigma_{SB|E}$ in Eq.~\eqref{eqn:state_e_block} can be indexed in two different ways: either in the same order as $\vb*{\epsilon}_{|E}$, written $\ket{\psi_{k|E}} = \ket{\epsilon_{k|E}}_S \ket{E-\epsilon_{k|E},\mu_{k|E}}_B$; or in the same order as $\vb{q}_{|E}$, denoted $\ket{\psi_{\pi_E(k)|E}}$ for the appropriate index permutation $\pi_E$.
The unitary then has the action
\begin{align}
    U_E \ket{\psi_{k|E}} = 
    \begin{cases}
        \frac{1}{\sqrt{2}} \left( \ket{\psi_{\pi_E^{-1}(k)|E}} + \ket{\psi_{\Delta_E - \pi_E^{-1}(k)|E}} \right) , & k < (\Delta_E-1)/2 \\
        \ket{\psi_{\pi_E^{-1}(k)|E}}, & k = (\Delta_E-1)/2 \\
        \frac{1}{\sqrt{2}} \left( \ket{\psi_{\pi_E^{-1}(k)|E}} - \ket{\psi_{\Delta_E - \pi_E^{-1}(k)|E}} \right) , & k > (\Delta_E-1)/2 
    \end{cases}.
\end{align}

\subsection{Low-temperature limit} 

\begin{figure}[h]
    \includegraphics[width=0.97\textwidth]{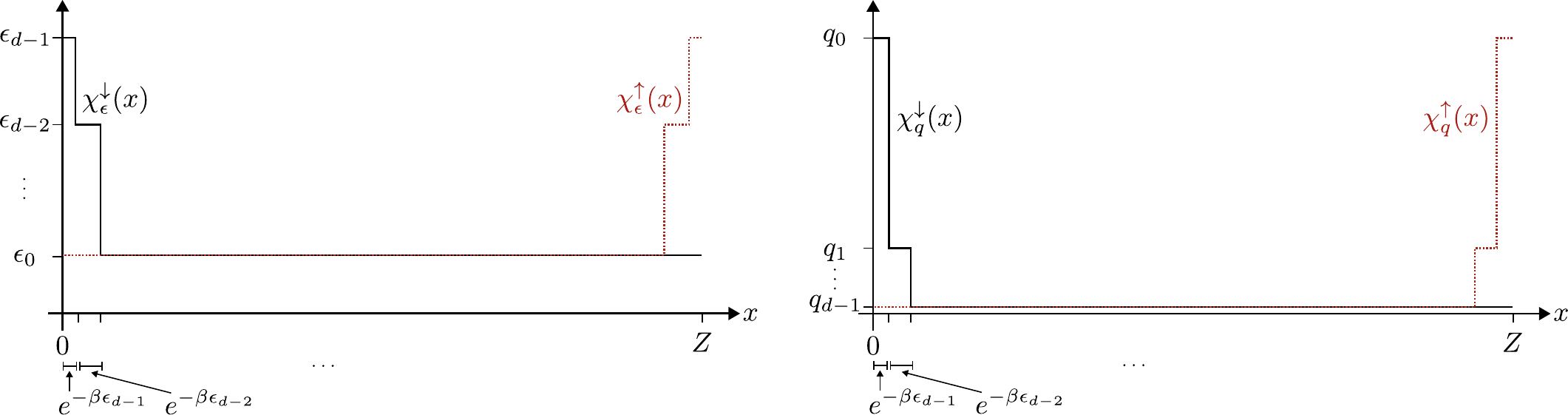}
    \caption{Illustration of step functions in the low-temperature limit. Here, $q_0 = e^{\beta \epsilon_{d-1}}p_{d-1}$, $q_1 = e^{\beta \epsilon_{d-2}}p_{d-2}$, and so on.  }
    \label{fig:steps_low_t}
\end{figure}

First assume that $\rho_S$ is full-rank, so that all $p_i > 0$.
As $T \to 0$, $\vb{q}^\downarrow$ and $\vb*{\epsilon}^\downarrow$ are in the same order and have the same pattern of repetitions.
For both $\chi_\epsilon^\downarrow$ and $\chi_q^\downarrow$, almost all the $x$-range is taken up by the lowest energy $\epsilon_0$ -- see Fig.~\ref{fig:steps_low_t} -- hence
\begin{align}
    \lim_{T \to 0} \, \mc{F}_T(\rho_S, H_S) & = \lim_{\beta \to \infty} \sum_{j=1}^{d-1} f(p_j e^{\beta \epsilon_j}, p_0 e^{\beta \epsilon_0}) (\epsilon_j - \epsilon_0)^2 \cdot e^{-\beta \epsilon_j} \nonumber \\
        & = \sum_{j=1}^{d-1} \lim_{\beta \to \infty} f(p_j, p_0 e^{\beta (\epsilon_0 - \epsilon_j)}) (\epsilon_j - \epsilon_0)^2 \nonumber \\
        & = \sum_{j=1}^{d-1} p_j (\epsilon_j - \epsilon_0)^2.
\end{align}
If any $p_i = 0$, the function $\chi_q^\downarrow$ is now ``missing'' some steps, so there is misalignment with $\chi_\epsilon^\downarrow$, and the sum is more complicated.
However, it is not too hard to see that, due to the step of width $e^{-\beta \epsilon_{d-2}}$ being exponentially wider than that of width $e^{-\beta \epsilon_{d-1}}$, and so on, each term of the form $p_j (\epsilon_k - \epsilon_0)^2$ has vanishing weight unless $k=j$, so the same result is recovered.

\subsection{High-temperature limit}
As $T \to \infty$, all Boltzmann factors tend to unity, $e^{-\beta \epsilon_i} \to 1$, so $\vb{q} = \vb{p}^\downarrow$.
In addition, the step lengths in $\chi_\epsilon^\downarrow$ and $\chi_q^\downarrow$ are all unity, so
\begin{align} \label{eqn:high_t}
    \lim_{T \to \infty} \, \mc{F}_T(\rho_S, H_S) & = \sum_{i=0}^{(d-1)/2} f(p^\downarrow_i, p^\uparrow_i) (\epsilon^\downarrow_i - \epsilon^\uparrow_i)^2.
\end{align}

\subsection{Qubit case} \label{app:discrete_qubit}
In the case $d=2$, we present a separate, simpler derivation of Eq.~\eqref{eqn:qfi_qubit} in which a bath with uniform degeneracies $D(k\epsilon) = 1 \; \forall \, k = 0,1,2, \dots$ is optimal. \\

Given $H_S = \epsilon \dyad{\epsilon}$, we have
\begin{align}
    \vb{q}_{|E} & = (p_0 \; [D(E) \text{ times}], \, p_1 e^{\beta \epsilon} \; [D(E-\epsilon) \text{ times}] \, )^\downarrow, \nonumber \\
    \vb*{\epsilon}_{|E} & = (0 \; [D(E) \text{ times}], \, \epsilon \; [D(E-\epsilon) \text{ times} \,] ).
\end{align}
The ordering of $\vb{q}_{|E}$ depends on comparing $p_0$ with $p_1 e^{\beta \epsilon}$.
Suppose first that $p_0 \geq p_1 e^{\beta \epsilon}$; then $\vb{q}_{|E}$ and $\vb*{\epsilon}_{|E}$ are ordered with the same pattern of repetitions.
The summand in Eq.~\eqref{eqn:qfi_e_block} vanishes when either $q_{k|E} = q_{\Delta_E-1-k|E}$ or $\epsilon_{\Delta_E-1-k|E} = \epsilon_{k|E}$, so that the number of nonzero terms is the minimum of $D(E)$ and $D(E-\epsilon)$.
Hence, we have
\begin{align}
    \max_{U_E} \, \mc{F}(P_E \, U_E \sigma_{SB|E} U_E^\dagger, H_S) = \frac{e^{-\beta E}}{Z_B} f(p_0, p_1 e^{\beta \epsilon}) \epsilon^2 \, \min \{D(E), D(E-\epsilon)\}.
\end{align}
In the other case that $p_0 < p_1 e^{\beta \epsilon}$, the result is in fact the same due to symmetry of the summand under reversing the order of $\vb{q}_{|E}$.
Overall, therefore,
\begin{align} \label{eqn:qubit_eqchain}
    \max_U \, \mc{F}(U \sigma_{SB} U^\dagger, H_S) & = f(p_0, p_1 e^{\beta \epsilon}) \epsilon^2 \sum_E \frac{e^{-\beta E}}{Z_B} \, \min \{D(E), D(E-\epsilon)\} \nonumber \\
        & \leq f(p_0, p_1 e^{\beta \epsilon}) \epsilon^2 \sum_E \frac{e^{-\beta E}}{Z_B} D(E-\epsilon) \nonumber \\
        & = f(p_0, p_1 e^{\beta \epsilon}) \epsilon^2 \sum_E \frac{e^{-\beta (E+\epsilon)}}{Z_B} D(E) \nonumber \\
        & = f(p_0, p_1 e^{\beta \epsilon}) \epsilon^2 e^{-\beta \epsilon} \sum_E \frac{e^{-\beta E}}{Z_B} D(E) \nonumber \\
        & = f(p_0 e^{-\beta \epsilon}, p_1) \epsilon^2.
\end{align}
The above inequality is saturated using degeneracies that satisfy $D(E-\epsilon) \leq D(E) \; \forall \, E$.
This is impossible for any finite-dimensional bath.
However, we show how saturation can be approached arbitrarily closely.
Let $E_{\max}$ be the lowest bath energy for which the condition fails -- i.e., for which $D(E_{\max} - \epsilon) > D(E_{\max})$.
Then we have
\begin{align}
    \sum_E \frac{e^{-\beta E}}{Z_B} \, \min \{D(E), D(E-\epsilon)\} & \geq \sum_{E < E_{\max}} \frac{e^{-\beta E}}{Z_B} D(E-\epsilon) \nonumber \\
    & = e^{-\beta \epsilon} \sum_{E < E_{\max} - \epsilon} \frac{e^{-\beta E}}{Z_B} D(E) \nonumber \\
    & = e^{-\beta \epsilon} P_B(E<E_{\max}-\epsilon),
\end{align}
where $P_B(E<E_{\max}-\epsilon)$ is the probability of the bath energy being less than $E_{\max} - \epsilon$, so
\begin{align}
    \max_U \, \mc{F}(U \sigma_{SB} U^\dagger, H_S) \geq f(p_0 e^{-\beta \epsilon}, p_1) \epsilon^2 P_B(E<E_{\max}-\epsilon).
\end{align}
The probability factor can be made arbitrarily close to 1 by constructing a bath with $E_{\max}$ sufficiently large. \\

\begin{figure}[h]
    \includegraphics[width=0.55\textwidth]{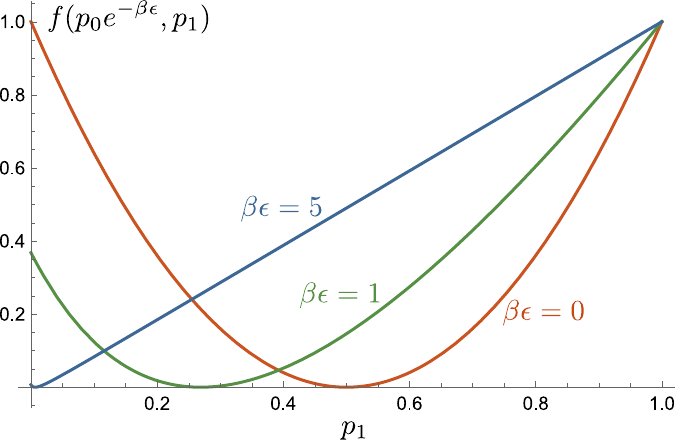}
    \caption{The maximum QFI $\mc{F}^\text{(qubit)}_T$ for a qubit with energy gap $\epsilon$ combined with a thermal bath at temperature $T$, in units of $\epsilon^2$ as a function of the excitation probability $p_1$, for three different temperatures.}
    \label{fig:qubit}
\end{figure}

So we can see that a bath with ladder Hamiltonian $H_B = \sum_{k=0}^K k \epsilon \dyad{k \epsilon}$ approaches the optimal qubit QFI as $K \to \infty$.
An alternative model is a single free particle in 1D, which has all energies $E \geq 0$ with uniform density of states.
The dependence of the optimal QFI on excitation probability $p_1$ and background temperature $T$ is shown in Fig.~\ref{fig:qubit}.

\subsection{Properties of $\mc{F}_T$}

Here, we characterise $\mc{F}_T$ as a formal quantifier of athermality due to the following properties:
\begin{enumerate}
    \item $\mc{F}_T(\rho_S,H_S) = 0$ if and only if $\rho_S = \gamma_S$.
    \item For any thermal operation $\Lambda$, we have $\mc{F}_T(\Lambda[\rho_S],H_S) \leq \mc{F}_T(\rho_S,H_S)$.
\end{enumerate}

For property (1), if $\rho_S = \gamma_S$, then $\gamma_S \ox \gamma_B \propto e^{-\beta(H_S+H_B)}$ is invariant under all energy-conserving unitaries, hence $\mc{F}_T$ vanishes.
Conversely, if $\rho_S \neq \gamma_S$, then there exists a pair $i \neq j$ such that $\epsilon_i \geq \epsilon_j$ and $p_i e^{\beta \epsilon_i} \neq p_j e^{\beta \epsilon_j}$.
We choose a bath with two nondegenerate energy levels at $0$ and $\epsilon_i - \epsilon_j$.
(Or, in the case $\epsilon_i = \epsilon_j$, we take a doubly degenerate level.)
The combined system-bath state projected onto energy $\epsilon_i$ is
\begin{align}
    \sigma_{SB|\epsilon_i} & \propto p_i \dyad{\epsilon_i}_S \ox \dyad{0}_B + p_j e^{-\beta (\epsilon_i-\epsilon_j)} \dyad{\epsilon_j}_S \ox \dyad{\epsilon_i-\epsilon_j}_B.
\end{align}
By assumption, this state is not maximally mixed, so by Lemma~\ref{lem:max_qfi_unitary} can be unitarily rotated to a state with nonzero QFI under $H_S$. \\

For property (2), we write the thermal operation as
\begin{align}
    \Lambda(\rho_S) & = \tr_{B'} [ V_{SB'} (\rho_S \ox \gamma_{B'}) V_{SB'}^\dagger],
\end{align}
where $[V_{SB'}, H_S + H_{B'}]=0$.
It follows that
\begin{align}
    \mc{F}_T(\Lambda[\rho_S],H_S) & = \sup_{H_B} \max_{U_{SB}} \, \mc{F}(U_{SB} [\Lambda\{\rho_S\}\ox \gamma_B] U_{SB}^\dagger, H_S) \nonumber \\
        & \leq \sup_{H_B} \max_{U_{SB}} \, \mc{F}(U_{SB} V_{SB'} [\rho_S \ox \gamma_B \ox \gamma_{B'}] V_{SB'}^\dagger U_{SB}^\dagger, H_S) \nonumber \\
        & \leq \mc{F}_T(\rho_S, H_S),
\end{align}
where the first inequality is due to the monotonicity of the QFI under partial trace over $B'$~\cite{Petz2002Covariance}, and the second follows from the pair of $H_B + H_B'$ and $U_{SB}V_{SB'}$ forming a restricted parameter range over which to optimise a combined bath $BB'$.\\

Viewing $\mc{F}_T(\rho_S,H_S)$ as a function of the probability distribution $\vb{p}$, a statement equivalent to (2) is that $\mc{F}_T$ respects the ordering described by \emph{thermo-majorisation}~\cite{Horodecki2013Fundamental,Ruch1976Principle}.
Standard majorisation~\cite{MarshallBookCh3} is defined in the following way: we say that a probability vector $\vb{x}$ is majorised by $\vb{y}$, denoted $\vb{x} \prec \vb{y}$, when
\begin{align}
    \sum_{i=0}^k x^\downarrow_i & \leq \sum_{i=0}^k y^\downarrow_i \quad \forall \, k = 0,1,\dots, d-2, \\
    \sum_{i=0}^{d-1} x^\downarrow_i & = \sum_{i=0}^{d-1} y^\downarrow_i .
\end{align}
This can be interpreted in terms of the Lorenz curve $L(\vb{x})$, defined as the piecewise linear function joining the points $(k, \sum_{i=0}^k x^\downarrow_i)$ -- i.e, a plot of the cumulative sums of $\vb{x}^\downarrow$, which is concave due to the ordering.
$\vb{x} \prec \vb{y}$ is equivalent to the statement that $L(\vb{x})$ lies (not strictly) below $L(\vb{y})$.

For thermo-majorisation, one instead $\beta$-orders $\vb{x}$ with an index permutation $\pi$, such that $x_{\pi(0)} e^{\beta \epsilon_{\pi(0)}} \geq x_{\pi(1)} e^{\beta \epsilon_{\pi(1)}} \geq \dots$ and constructs the curve $T(\vb{x})$ by joining the points $(\sum_{i=0}^k e^{-\beta \epsilon_{\pi(i)}}, \sum_{i=0}^k x_{\pi(i)})$.
We say that $\vb{x}$ is thermo-majorised by $\vb{y}$, denoted $\vb{x} \prec_\gamma \vb{y}$, when $T(\vb{x})$ lies (not strictly) below $T(\vb{y})$.
Given a pair of states diagonal states $\rho_S = \sum_i r_i \dyad{\epsilon_i}$, $\sigma_S = \sum_i s_i \dyad{\epsilon_i}$, the existence of a thermal operation $\Lambda$ mapping between them, $\Lambda(\rho_S) = \sigma_S$, is equivalent to $\vb{r} \succ_\gamma \vb{s}$~\cite{Horodecki2013Fundamental}.
(This reduces to standard majorisation when $\gamma$ is maximally mixed, i.e., in the high-temperature limit.)

Therefore, $\mc{F}_T$ respects thermo-majorisation ordering in the sense that
\begin{align}
    \vb{r} \succ_\gamma \vb{s}\; \Longrightarrow \; \mc{F}_T\left(\sum_i r_i \dyad{\epsilon_i}, H_S\right) \geq \mc{F}_T\left(\sum_i s_i \dyad{\epsilon_i}, H_S\right).
\end{align}

\section{Large bath lemma} \label{app:large_bath}

In this section, we prove the following claim:
\begin{lem} \label{lem:large_bath}
    \begin{mdframed}
        For any finite bath $B$ with degeneracies $D(E)$, there exists another bath $B'$ such that the combination $\tilde{B} = BB'$ with $H_{\tilde{B}} = H_B + H_{B'}$ has degeneracies satisfying $\tilde{D}(E+\epsilon) \approx \tilde{D}(E)e^{\beta \epsilon}$ to arbitrarily small fractional error for all system energies $\epsilon$, with arbitrarily high probability when $E$ is sampled from the energies of $\tilde{B}$.
        Furthermore, $\tilde{B}$ can be chosen to have an arbitrarily dense energy spectrum.
    \end{mdframed}
\end{lem}

We start with an arbitrary intial finite bath $B$, and suppose we append an additional ideal infinite bath $B'$ that satisfies $D'(E+\epsilon) = D'(E) e^{\beta \epsilon}$ exactly.
This property is retained when $B'$ is appended to the original bath $B$: the combined degeneracy $\tilde D$ of $\tilde{B}=BB'$ is given by a convolution
\begin{align}
    \tilde D(E) = \sum_\eta \; D(\eta) D'(E-\eta)
\end{align}
and so satisfies
\begin{align} \label{eqn:exp_scaling_stable}
    \tilde D(E+\epsilon) & = \sum_\eta \; D(\eta) D'(E-\eta+\epsilon) \nonumber \\
        & = \sum_\eta \; D(\eta) D'(E-\eta) e^{\beta \epsilon} \nonumber \\
        & = \tilde D(E) e^{\beta \epsilon}.
\end{align}

Exact exponential scaling for all energies is however unphysical -- for instance, it leads to a diverging partition function and implies an infinite heat capacity~\cite{Lostaglio2019Introductory,Masanes2017Derivation,Richens2018Finite}.
Rather, a realistic large bath will have a certain interval of energies for which exponential scaling holds approximately.
Formally, we take some arbitrarily small error $\delta > 0$ and demand that there be an interval $I_\delta$ such that
\begin{align} \label{eqn:exp_scaling_approx}
    \abs{\frac{D'(E) e^{\beta \epsilon}}{D'(E+\epsilon)} - 1} \leq \delta \quad \forall \; E, \, E+\epsilon \in I_\delta.
\end{align}
The approximate version of Eq.~\eqref{eqn:exp_scaling_stable} becomes
\begin{align}
    \abs{ \tilde D(E+\epsilon) - \tilde D(E) e^{\beta \epsilon}} & = \abs{ \sum_\eta \; D(\eta) \left[ D'(E-\eta+\epsilon) - D'(E-\eta)e^{\beta \epsilon} \right] } \nonumber \\
        & \leq \sum_\eta \; D(\eta) \abs{ D'(E-\eta+\epsilon) - D'(E-\eta)e^{\beta \epsilon} } \nonumber \\
        & \leq \sum_\eta \; D(\eta) D'(E-\eta+\epsilon) \delta \nonumber \\
        & = \tilde D(E+\epsilon) \delta,
\end{align}
as long as $E - \eta ,\, E - \eta + \epsilon \in I_\delta$ for all $\eta$ where $D(\eta)$ is nonzero.\\

\begin{figure}[h]
    \includegraphics[width=.6\textwidth]{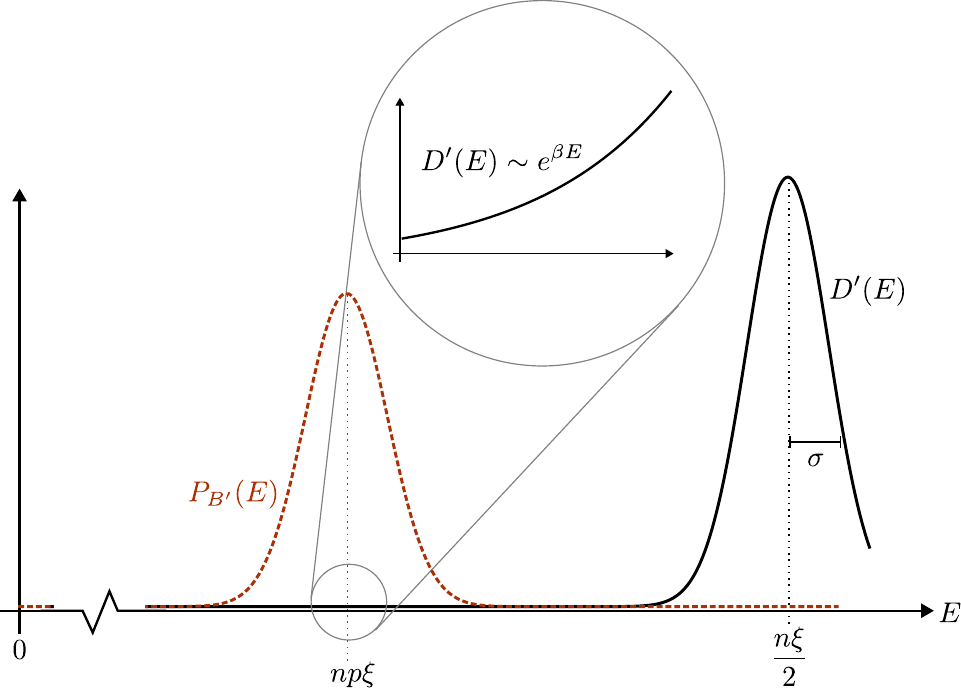}
    \caption{The large bath $B'$ constructed in this section has energies $E = k\xi$ for $k = 0,1,\dots,n$, with $\xi$ being small so that the degeneracy $D'(E)$ and probability distrubtion $P_{B'}(E)$ are depicted as continuous. Both distributions are peaked around their means, each being a finite fraction of $n\xi$, with much smaller width $\sigma \sim \sqrt{n}\xi$. Within the peak of most probable energies, the degeneracy scales close to exponentially in energy.}
    \label{fig:large_bath}
\end{figure}

The second requirement must then be that the probability of having eneriges in this interval is arbitrarily close to 1.
It is easy to construct such a $B'$.
For example, we take $n$ identical noninteracting qubits, each with energy gap $\xi$, chosen small enough that all gaps in the system's energy spectrum are well approximated within the gaps of $H_{B'}$.
This is done to effectively treat the bath as a continuum, so that all system energies are accounted for in the state decomposition \eqref{eqn:state_e_block}.
(See Ref.~\cite{vomEnde2022Which} for a discussion of issues related to fine-tuning the bath spectrum and continuity of the set of thermal operations.)
We further assume $\xi$ to be small enough that $\beta \xi \ll 1$.
It follows that each qubit is excited with probability $p = e^{-\beta\xi} / (1 + e^{-\beta\xi}) \approx 1/2 - \beta\xi/4$.
If the number of excited qubits is $k$, then the law of large numbers says that, for sufficiently large $n$,
\begin{align} \label{eqn:typical_k}
    \abs{\frac{k}{n} - p} < \Delta
\end{align}
with arbitrarily high probability, where $\Delta > 0$ is some small arbitrary threshold (which will later be related to $\delta$).
The interval of typical energies is therefore chosen as $I_\delta = [n(p-\Delta) \xi, n(p+\Delta) \xi]$.

The degeneracy at energy $E=k\xi$ is
\begin{align}
    D'(k\xi) & = \binom{n}{k},
\end{align}
and we are interested in the ratio $D'([k+l]\xi) / D'(k\xi)$ for any typical $k$ according to Eq.~\eqref{eqn:typical_k} and $l \ll n$.
We compute
\begin{align}
    \ln \left[ \frac{D'([k+l]\xi)}{D'(k\xi)} \right] & = \ln \binom{n}{k+l} - \ln \binom{n}{k} \nonumber \\
        & = \ln \left[ \frac{(n-k)(n-k-1) \dots (n-k-l+1)}{(k+l)(k+l-1) \dots (k+1)} \right] \nonumber \\
        & = \left[ \sum_{j=0}^{l-1} \ln \left(1 - q + \frac{j}{n} \right) \right] - \left[ \sum_{j=1}^l \ln \left(q + \frac{j}{n}\right) \right] \nonumber \\
        & = l \ln \left(\frac{1-q}{q}\right) + \mc{O}(n^{-1}),
\end{align}
where $q = k/n$.
Given $\abs{q-p} < \Delta$, we also have (from a first-order Taylor expansion)
\begin{align}
    \ln \left(\frac{1-q}{q}\right) & = \ln \left(\frac{1-p}{p}\right) + (q-p) \left(\frac{1}{1-p} - \frac{1}{p}\right) + \mc{O}(\Delta^2) \nonumber \\
        & = \beta \xi + (q-p) (1+e^{-\beta\xi})(1-e^{\beta\xi}) + \mc{O}(\Delta^2) \nonumber \\
        & = \beta \xi + \mc{O}( \Delta [\beta\xi]^2 ) + \mc{O}(\Delta^2) .
\end{align}
So with probability arbitrarily close to 1, the energy $E$ of $B'$ is in $I_\delta$, and given $\epsilon / \xi = l \ll n$, we have
\begin{align}
    \ln \left[ \frac{D'(E+\epsilon)}{D'(E)} \right] & = \beta \epsilon + \mc{O}(n^{-1}) + \mc{O}( \Delta [\beta\xi]^2 ) + \mc{O}(\Delta^2) , \nonumber \\
    \frac{D'(E)e^{\beta \epsilon}}{D'(E+\epsilon)} & = e^{\mc{O}(n^{-1}) + \mc{O}( \Delta [\beta\xi]^2 ) + \mc{O}(\Delta^2) } \nonumber \\
        & = 1 + \mc{O}(n^{-1}) + \mc{O}( \Delta [\beta\xi]^2 ) + \mc{O}(\Delta^2) .
\end{align}
Therefore, Eq.~\eqref{eqn:exp_scaling_approx} is satisfied with $\delta = \mc{O}(n^{-1}) + \mc{O}( \Delta [\beta\xi]^2 ) + \mc{O}(\Delta^2) $.
In order to guarantee $\epsilon / \xi = l \ll n$, we just need to choose $n$ sufficiently large compared with $\epsilon/\xi$ for any energy $\epsilon$ of the system $S$ and original bath $B$ combined.
Fig.~\ref{fig:large_bath} illustrates the probability and degeneracy of $B'$ as a function of energy.

\section{Finite-dimensional interaction speed} \label{app:speed_discrete}
\begin{thm}
    \begin{mdframed}
        The maximum interaction speed of a finite-dimensional system with a thermal bath is
        \begin{align}
            \mc{F}_T^\text{int}(\rho_S,H_S) & := \sup_{H_B} \, \max_{\substack{H_I \\ [H_I, H_S+H_B]=0, \\ \spr(H_{I|E}) \leq 1 \,\ \forall E}} \mc{F}(\rho_S \ox \gamma_B, H_I) \nonumber \\
                & = \int_0^{Z_S/2} \dd x \; f(\chi_q^\downarrow[x], \chi_q^\uparrow[x]).
        \end{align}
    \end{mdframed}
\end{thm}

For the maximum interaction speed, we follow the same approach leading to Eq.~\eqref{eqn:qfi_e_block}, except that now we need to use the generator $H_I = \bigoplus_E H_{I|E}$.
The projection $H_{I|E}$ of this onto total energy $E$ has eigenvalues $h_{k|E}$, so from Lemma~\ref{lem:max_qfi_unitary},
\begin{align}
    \max_{H_{I|E}} \, \mc{F}(P_E \, \sigma_{SB|E}, H_{I|E}) = \frac{e^{-\beta E}}{Z_B} \sum_{k=0}^{(\Delta_E-1)/2} f(q_{k|E}, q_{\Delta_E-1-k|E}) (h_{\Delta_E-1-k|E} - h_{k|E})^2.
\end{align}
Due to the imposed constraint $\spr(H_{I|E}) \leq 1$, we have $\abs{h_{j|E} - h_{k|E}} \leq 1 \; \forall \, j,k$, so
\begin{align} \label{eqn:interaction_bound}
    \max_{H_{I|E}} \, \mc{F}(P_E \, \sigma_{SB|E}, H_{I|E}) \leq \frac{e^{-\beta E}}{Z_B} \sum_{k=0}^{(\Delta_E-1)/2} f(q_{k|E}, q_{\Delta_E-1-k|E}).
\end{align}
Following the same line of argument as in Appendix~\ref{app:discrete_proof}, the optimal bath has exponentially scaling degeneracies, leading us to construct the same step function $\chi_q^\downarrow(x)$.
However, we are now also free to vary the generator eigenvalues.
This implies the ability to choose any nonincreasing function $\chi_h^\downarrow$ whose maximum and minimum values differ by at most 1 -- we set the range to $[0,1]$ without loss of generality.
The upper bound Eq.~\eqref{eqn:interaction_bound} for this bath then becomes
\begin{align} \label{eqn:interaction_e_block}
    \max_{H_{I|E}} \, \mc{F}(P_E \, \sigma_{SB|E}, H_{I|E}) & = \frac{e^{-\beta E}}{Z_B} D(E) \, \max_{\chi_h^\downarrow} \int_0^{Z_S/2} \dd x \; f(\chi_q^\downarrow[x], \chi_q^\uparrow[x]) (\chi_h^\downarrow[x] - \chi_h^\uparrow[x])^2 \nonumber \\
        & \leq \frac{e^{-\beta E}}{Z_B} D(E) \int_0^{Z_S/2} \dd x \; f(\chi_q^\downarrow[x], \chi_q^\uparrow[x]).
\end{align}
One then sees that this inequality is saturated with $\chi_h^\downarrow(x) = 1$ for $x \in [0,Z_S/2]$ and otherwise 0.
Overall, therefore,
\begin{align}
    \mc{F}_T^\text{int}(\rho_S,H_S) =  \int_0^{Z_S/2} \dd x \; f(\chi_q^\downarrow[x], \chi_q^\uparrow[x]).
\end{align}
\qed \\

One may want to extend this result by relaxing the spectrum constraint on $H_I$ to be a chosen function of $E$, say $\spr(H_{I|E}) \leq s(E)$.
By rescaling $H_{I|E}$, the right-hand side of Eq.~\eqref{eqn:interaction_e_block} gets multiplied by $s(E)^2$, and the overall upper bound gains a factor of
\begin{align}
    \ev{s(E)^2} := \sum_E \frac{e^{-\beta E}}{Z_B} D(E) s(E)^2.
\end{align}
This factor depends on the choice of bath, and may be made arbitrarily large depending on the chosen asymptotic behaviour of $s(E)$.
However, it can be useful in some physical examples.
For instance, if $S$ is a qubit and $B$ is a single bosonic mode, then we can consider the Jaynes-Cummings Hamiltonian~\cite{Jaynes1963Comparison} with terms
\begin{align}
    H_S = \epsilon \dyad{\epsilon} , \quad H_B = \epsilon (a^\dagger a + 1/2), \quad H_I = \frac{\Omega}{2} (\dyad{0}{\epsilon} \ox a^\dagger + \dyad{\epsilon}{0} \ox a).
\end{align}
The interaction decomposes into blocks
\begin{align}
    H_{I|n\epsilon} = \frac{\Omega \sqrt{n}}{2} \left( \dyad{0}{\epsilon} \ox \dyad{n}{n-1} + \dyad{\epsilon}{0} \ox \dyad{n-1}{n} \right),
\end{align}
thus $s(n\epsilon) = \Omega \sqrt{n}$.
From Appendix~\ref{app:discrete_qubit}, we know that $B$ has optimal degeneracies saturating the upper-bound in Eq.~\eqref{eqn:qubit_eqchain}, thus the QFI under $H_I$ is
\begin{align}
    \sum_{n=1}^\infty f(p_0, p_1 e^{\beta \epsilon}) \frac{e^{-\beta \epsilon n}}{Z_B} s(n\epsilon)^2 & = \sum_{n=1}^\infty f(p_0, p_1 e^{\beta \epsilon}) \frac{e^{-\beta \epsilon n}}{Z_B} \Omega^2 n \nonumber \\
        & = \Omega^2 f(p_0, p_1 e^{\beta \epsilon}) \cdot \frac{1}{e^{\beta \epsilon}-1} \nonumber \\
        & = \Omega^2 \frac{f(p_0 e^{-\beta \epsilon}, p_1 )}{1-e^{-\beta \epsilon}}.
\end{align}
This has low- and high-temperature limits $\Omega^2 p_1$ and $\Omega^2 (2p_1-1)^2 \kb T / \epsilon$, respectively.

\section{Optical setting: Derivation of latent coherence} \label{app:max_qfi_optical}
Here, we prove the main result in the optical setting:
\begin{thm} \label{thm:max_qfi_optics}
    \begin{mdframed}
        For any single-mode state $\rho_S = \sum_{n=0}^\infty p_n \dyad{n}$, the maximum QFI for sensing a phase generated by $H_S = \hbar \omega N_S$, using any number of ancilliary thermal modes plus linear optics, is
        \begin{align}
            \sup_k \max_{U \in \mathrm{LO}} \mc{F}(U[\rho_S \ox \gamma^{\ox k}_B]U^\dagger, H_S) & = (\hbar \omega)^2 (\nth + 1) \coh_r(\rho_S), \\
            \coh_r(\rho_S) & = \sum_{n=1}^\infty f(p_n, r p_{n-1}) n,
        \end{align}
        with the supremum achieved for $k=1$ and $U$ being a single balanced beam splitter.
    \end{mdframed}
\end{thm}

A joint passive linear unitary on all $(k+1)$ modes can actually always be reduced to a single beam splitter with one ancilla mode $B$~\cite{Narasimhachar2021Thermodynamic,Serafini2020Gaussian} -- this is due to the cosine-sine decomposition of such unitaries~\cite[Appendix B]{Yadin2018Operational} and the invariance of $\gamma_B^{\ox k}$ under all $k$-mode passive linear unitaries.
So we only need consider an unbalanced beam splitter with the action $U^\dagger a U = ua + vb,\, \abs{u}^2 + \abs{v}^2 = 1$.
We have
\begin{align}
    U^\dagger N_S U & = (u^* a^\dagger + v^* b^\dagger)(u a + v b) \nonumber \\
        & = \abs{u}^2 a^\dagger a + \abs{v}^2 b^\dagger b + u^* v a^\dagger b + v^* u b^\dagger a \nonumber \\
        & = \abs{u}^2 N_S + \abs{v}^2 N_B + u^* v a^\dagger b + v^* u b^\dagger a .
\end{align}
The QFI is
\begin{align}
    \mc{F}(U[\rho_S \ox \gamma_B]U^\dagger, N_S) & = \mc{F}(\rho_S \ox \gamma_B, U^\dagger N_S U) \nonumber \\
        & = \mc{F}(\rho_S \ox \gamma_B, \abs{u}^2 N_S + \abs{v}^2 N_B + u^* v a^\dagger b + v^* u b^\dagger a) \nonumber \\
        & = \mc{F}(\rho_S \ox \gamma_B, \abs{u v} [ a^\dagger b +b^\dagger a ]) \nonumber \\
        & = \abs{uv}^2 \mc{F}(\rho_S \ox \gamma_B, a^\dagger b +b^\dagger a ),
\end{align}
where the third line follows from phase-invariance of the input states, which removes contributions from the number operator generators and lets us cancel the phase of $u^* v$ through $a \to e^{i \varphi} a$.
The maximum value of $\abs{uv}^2$ is $1/4$, attained for a balanced beam splitter with $u = v = 1/\sqrt{2}$, hence
\begin{align} \label{eqn:optical_1bs}
    \max_{U \in \mathrm{LO}} \, \mc{F}(U[\rho_S \ox \gamma_B]U^\dagger, N_S) & = \frac{1}{4} \mc{F}(\rho_S \ox \gamma_B, G), \nonumber \\
    G & := a^\dagger b +b^\dagger a .
\end{align}

From this point, we simply evaluate the right-hand side of Eq.~\eqref{eqn:optical_1bs} using the spectral decomposition of the joint state
\begin{align}
    \rho_S \ox \gamma_B & = \left( \sum_{n=0}^\infty p_n \dyad{n}_S  \right) \ox \left( \sum_{m=0}^\infty g_m \dyad{m}_B \right) \nonumber \\
        & = \sum_{n,m=0}^\infty p_n g_m \dyad{n,m},
\end{align}
where $g_m = (1-r) r^m$, $r = e^{-\beta \hbar \omega}$.
From the expression in Eq.~\eqref{eqn:qfi_expression} giving
\begin{align}
    \mc{F}(\rho_S \ox \gamma_B, G) = 2 \sum_{n,m,n',m'} f(p_n g_m, p_{n'} g_{m'}) \abs{\mel{n, m}{G}{n', m'}}^2
\end{align}
and the matrix element
\begin{align}
    \mel{n, m}{G}{n', m'} & = \mel{n}{a^\dagger}{n'} \mel{m}{b}{m'} + \mel{n}{a}{n'} \mel{m}{b^\dagger}{m'} \nonumber \\
        & = \delta_{n',n-1} \delta_{m',m+1} \sqrt{n(m+1)} + \delta_{n',n+1} \delta_{m',m-1} \sqrt{(n+1)m},
\end{align}
we obtain
\begin{align}
    \mc{F}(\rho_S \ox \gamma_B, G) & = 2\sum_{n,m} f(p_n g_m, p_{n-1} g_{m+1}) n(m+1) + f(p_n g_m, p_{n+1} g_{m-1}) (n+1)m
    \nonumber \\
        & = 2 \sum_{n,m} f(p_n g_m, p_{n-1} g_{m+1}) n(m+1) + f(p_{n-1} g_{m+1}, p_n g_m) n(m+1) \nonumber \\
        & = 4\sum_{n,m} f(p_n g_m, p_{n-1} g_{m+1}) n(m+1) \nonumber \\
        & = 4 \sum_{n,m} f(p_n, p_{n-1} g_{m+1}/g_m) g_m n(m+1),
\end{align}
having used properties (1) and (2) from Proposition~\ref{prop:fisher_diff}.
Finally, we use the property of the thermal distribution $g_{m+1}/g_m = r$ and then the sum factorises:
\begin{align}
    \mc{F}(\rho_S \ox \gamma_B, G) & = 4 \left[ \sum_m g_m (m+1) \right] \left[ \sum_n f(p_n, rp_{n-1}) n \right] \nonumber \\
        & = 4 (\nth + 1) \coh_r(\rho_S).
\end{align}\\

Due to Proposition~\ref{prop:fisher_diff}, we can make the following observations:
\begin{itemize}
    \item $\coh_r$ is convex, i.e., $\coh_r(p \rho + [1-p] \sigma) \leq p \coh_r(\rho) + (1-p)\coh_r(\sigma)$ for $p \in [0,1]$. Note that $\coh_0$ is actually linear.
    \item $\coh_r$ is guaranteed to be (locally) increasing in $T$ if $p_n / p_{n-1} < r \; \forall \, n \geq 1$ (and decreasing in $T$ for the reverse inequality). This condition means that $\rho$ is unambiguously colder than its environment, in the sense that the occupations of neighbouring energy levels all have a ratio smaller than the Boltzmann factor. This can be seen from property (5), considering the derivative $\partial_r f(p_n, r p_{n-1})$ and the fact that $\partial_T r > 0$.
    \item The general upper bound holds:
    \begin{align} \label{eqn:coh_upper}
        \coh_r(\rho) \leq (1+r) \ev{N}_\rho + r = \frac{(2\nth +1)\ev{N}_\rho + \nth}{\nth + 1},
    \end{align}
    which is attained (if $r>0$) for any state with only even or odd photon numbers, equivalently $\tr[\rho a \rho a^\dagger] = 0$. This is seen from the trade-off result in Theorem~\ref{thm:tradeoff}.
\end{itemize}

A measurement $M$, for which the classical Fisher information always attains the upper limit of the QFI, can be expressed formally in terms of the symmetric logarithmic derivative~\cite{Braunstein1994Statistical}.
For our optical setting, this is
\begin{align}
    M & = \frac{2}{\mc{F}(\rho_S \ox \gamma_B, G)} \sum_{n,m,n',m'} \frac{p_n g_m - p_{n'}g_{m'}}{p_n g_m + p_{n'} g_{m'}} \mel{n,\, m}{-i G}{n',\, m'} \ketbra{n,\, m}{n',\, m'} \nonumber \\
        & = \frac{-2i}{(\nth+1)\coh_r(\rho_S)} \sum_{n,m} \frac{p_n g_m - p_{n-1} g_{m+1}}{p_n g_m + p_{n-1} g_{m+1}} \sqrt{n(m+1)} \left( \ketbra{n,\, m}{n-1,\, m+1} - \ketbra{n-1,\, m+1}{n,\, m} \right) \nonumber \\
        & = \frac{-2i}{(\nth+1)\coh_r(\rho_S)} \sum_{n,m} \frac{p_n - r p_{n-1}}{p_n + r p_{n-1}} \sqrt{n(m+1)} \left( \ketbra{n,\, m}{n-1,\, m+1} - \ketbra{n-1,\, m+1}{n,\, m} \right) \nonumber \\
        & = \frac{-2i}{(\nth+1)\coh_r(\rho_S)} \sum_n  \frac{p_n - r p_{n-1}}{p_n + r p_{n-1}} \sqrt{n} \left( \ketbra{n}{n-1} b - \ketbra{n-1}{n}b^\dagger \right) \nonumber \\
        & = \frac{-2i}{(\nth+1)\coh_r(\rho_S)} \left( \sum_n  \frac{p_n - r p_{n-1}}{p_n + r p_{n-1}} \sqrt{n} \ketbra{n}{n-1} \right) b + \text{h.c.} \nonumber \\
        & \propto -i \left( \tilde{a}^\dagger b - \tilde{a} b^\dagger \right),
\end{align}
where
\begin{align}
    \tilde{a} &:= \sum_n  \frac{p_n - r p_{n-1}}{p_n + r p_{n-1}} \sqrt{n} \ketbra{n-1}{n}
\end{align}
is a version of the annihilation operator $a$ with reweighted matrix elements.
For $r = 0$, $\tilde{a} = a$, recovering the standard homodyne detection strategy.
For $r = 1$, $\tilde{a}$ corresponds to the optimal observable for sensing quadrature displacements.
When $\rho = \gamma'$ is a thermal state at any temperature $T'$, the homodyne strategy appears again: given $p_n = r' p_{n-1}$, we have
\begin{align}
    \tilde{a} & = \sum_n \frac{(r'-r) p_{n-1}}{(r'+r)p_{n-1}} \sqrt{n} \dyad{n-1}{n} \nonumber \\
        & = \left( \frac{r'-r}{r'+r} \right) a.
\end{align}

\section{Optical state examples} \label{app:optics_examples}
\subsection{Fock state}
For a single nonvanishing $p_n$, only two terms remain in the sum: one pairing $p_n = 1$ with each of $p_{n-1} = 0$ and $p_{n+1} = 0$, so
\begin{align}
    \coh_r(\dyad{n}) = f(1,0)n + f(0,r)(n+1) = (1+r)n + r.
\end{align}

\subsection{Thermal state}
We take $\rho = \gamma'$, thermal at temperature $T' \neq T$.
Since $p_n/p_{n-1} = r'$, we have
\begin{align} \label{eqn:thermal_state_coherence}
    \coh_r(\gamma') & = \sum_n f(r' p_{n-1}, r p_{n-1}) n \nonumber \\
        & = \sum_n f(r',r) p_{n-1} n \nonumber \\
        & = f(r',r) \ev{N+1}_{\gamma'} \nonumber \\
        & = \frac{(r'-r)^2}{(r'+r)(1-r')}.
\end{align}

\subsection{Poisson statistics}
Let $p_n = \frac{e^{-\lambda}\lambda^n}{n!}$ (which can be obtained from a phase-averaged coherent state $\ket{\alpha}$ with $\abs{\alpha}^2 = \lambda$).
We claim the following integral representation:
\begin{align} \label{eqn:poisson}
    \coh_r(\rho_\lambda) = r(\lambda+1) - 3\lambda + 4 \frac{\lambda}{r} - 4\frac{\lambda^2}{r^2} \int_0^1 \dd u \, u^{\frac{\lambda}{r}-1} e^{\lambda(u-1)}.
\end{align}
The large-amplitude limit at fixed temperature is
\begin{align}
    \lambda \gg 1 \Rightarrow \coh_r(\rho_\lambda) = \lambda \cdot \frac{(1-r)^2}{1+r} + r \left[1 - \frac{4}{(1+r)^3}\right] + \mc{O}(\lambda^{-1}). 
\end{align}
In particular, $\coh_1(\rho_\lambda) \approx \frac{1}{2}$ for $\lambda \gg 1$, which shows that phase-averaging a large-amplitude coherent state reduces its QFI under displacements by a factor of 2, since $\mc{F}(\dyad{\alpha},x) = 2$ while $\mc{F}(\rho_\lambda,x) = 2\coh_1(\rho_\lambda) \approx 1$.\\

\inlineheading{Proof of Eq.~\eqref{eqn:poisson}}
Using $f(x,y) = x + y - 4xy/(x+y)$, we can write
\begin{align}
    \coh_r(\rho_\lambda) & = \sum_n \left(p_n + rp_{n-1}  - \frac{4rp_n p_{n-1}}{p_n + rp_{n-1}}\right) n \nonumber \\
        & = \lambda + r(\lambda + 1) - 4 \sum_n \frac{r p_n p_{n-1}}{p_n + rp_{n-1}} n.
\end{align}
We turn this latter sum into an integral that can be evaluated numerically.
The Poisson distribution has the recurrence relation $p_n/p_{n-1} = \lambda/n$, so
\begin{align}
    \sum_n \frac{r p_n p_{n-1} n}{p_n + rp_{n-1}} & = \sum_n p_{n-1} \frac{r n}{1 + rn/\lambda} \nonumber \\
        & = \sum_n p_{n-1} \frac{\lambda n}{\lambda/r + n} \nonumber \\
        & = \lambda \sum_n p_{n-1} \left[1 - \frac{\lambda/r}{\lambda/r + n} \right] \nonumber \\
        & = \lambda - \frac{\lambda^2}{r} \sum_n p_n \frac{n}{\lambda} \cdot \frac{1}{\lambda/r+n} \nonumber \\
        & = \lambda - \frac{\lambda}{r} \sum_n p_n \left[1 - \frac{\lambda/r}{\lambda/r+n} \right] \nonumber \\
        & = \lambda - \frac{\lambda}{r} + \frac{\lambda^2}{r^2} \sum_n \frac{p_n}{\lambda/r+n}.
\end{align}
Now consider the integral
\begin{align}
    I(\lambda,r) := \int_0^1 \dd u\; u^{\frac{\lambda}{r}-1} e^{\lambda (u-1)},
\end{align}
which converges when $\lambda/r>0$.
By a series expansion, we have
\begin{align}
    I(\lambda,r) & = \sum_{n=0}^\infty e^{-\lambda}\int_0^1 \dd u\; u^{\frac{\lambda}{r}-1} \frac{\lambda^n u^n}{n!} \nonumber \\
        & = \sum_n e^{-\lambda} \frac{\lambda^n}{n!} \int_0^1 \dd u \; u^{\frac{\lambda}{r} - 1 + n} \nonumber \\
        & = \sum_n \frac{p_n}{\lambda/r + n} \nonumber.
\end{align}
Hence,
\begin{align}
    \coh_r(\rho_\lambda) & = \lambda + r(\lambda+1) -4\lambda + 4\frac{\lambda}{r} - 4\frac{\lambda^2}{r^2} I(\lambda,r) \nonumber \\
        & = r(\lambda+1) - 3\lambda + 4\frac{\lambda}{r}- 4\frac{\lambda^2}{r^2} I(\lambda,r).
\end{align} \\

To find the limit $\lambda \gg 1$, we rewrite the integral as
\begin{align}
    I(\lambda,r) & = \frac{1}{x-1} \int_0^{\lambda/r-1} \dd s\; \left(1-\frac{s}{\lambda/r-1}\right)^{\lambda/r-1} e^{-\frac{\lambda}{\lambda/r-1}s} \nonumber \\
        & = \frac{1}{x-1} \int_0^{x-1} \dd s\; \left(1 - \frac{s}{x-1}\right)^{x-1} e^{-\frac{\lambda}{x-1}s}
\end{align}
by changing variables to $s = (\lambda/r-1)(1-u)$ and letting $x = \lambda/r$.
With $x \geq \lambda \gg 1$, we can approximate this integral by
\begin{align}
    I(\lambda,r) & \approx \frac{1}{x-1} \int_0^\infty \dd s\; \left(1 - \frac{s^2}{2x}\right) e^{-s} e^{-\frac{\lambda}{x-1}s} \nonumber \\
        & = \frac{1}{x-1} \int_0^\infty \dd s\; \left(1 - \frac{s^2}{2x}\right) e^{-Ks},
\end{align}
with $K = 1 + \lambda/(x-1) = (x+\lambda-1)/(x-1)$.
This can be evaluated to give
\begin{align}
    I(\lambda,r) & \approx \frac{1}{(x-1)K} \left( 1 - \frac{1}{xK^2} \right),
\end{align}
and by expanding to second order in $1/\lambda$, we obtain
\begin{align}
    I(\lambda,r) & \approx \frac{r^2}{1+r} \cdot \frac{1}{\lambda} + \frac{r^2}{(1+r)^3} \cdot \frac{1}{\lambda^2}
\end{align}
and consequently
\begin{align}
    \coh_r(\rho_\lambda) & \approx r \left[1 - \frac{4}{(1+r)^3} \right] + \lambda \left[r - 3 + \frac{4}{1+r} \right] \nonumber \\
        & = r \left[1 - \frac{4}{(1+r)^3} \right] + \lambda \frac{(1-r)^2}{1+r} \nonumber \\
        & = \frac{\nth}{\nth+1} \left[1 - 4\left(\frac{\nth+1}{2\nth+1}\right)^3 \right] + \frac{\lambda}{(\nth+1)(2\nth+1)} .
\end{align}

\begin{figure}[h]
    \includegraphics[width=.98\textwidth]{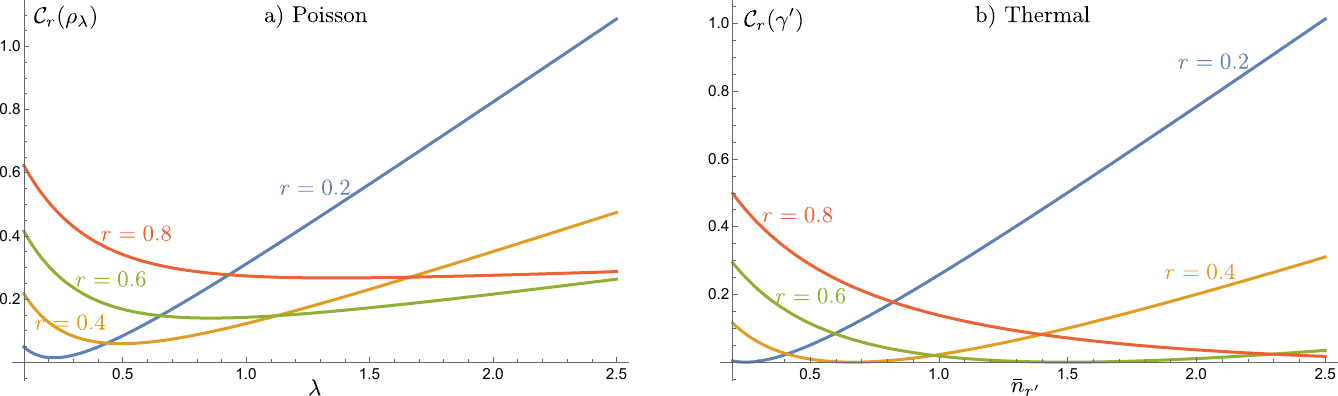}
    \caption{Latent coherence relative to different background temperatures, indexed by the Boltzmann ratio $r = e^{-\beta \hbar \omega}$, for two families of states, as a function of mean photon number. a) Poisson statistics $p_n = e^{-\lambda} \lambda^n / n!$ with mean $\lambda$; b) thermal statistics $p_n = (1-r') {r'}^n$ at temperature $T'$ with mean $\bar{n}_{r'} = r'/(1-r')$.}
    \label{fig:plots}
\end{figure}

\section{Nonclassicality witness}
Under the Glauber-Sudarshan definition of nonclassicality in optics, a state $\rho_\text{cl}$ is classical whenever it can be expressed as a convex combination of coherent states~\cite{Glauber1963Quantum,Sudarshan1963Equivalence,Mandel1986Non-classical}.
For $m$ modes, a coherent state is a product $\ket{\vb*{\alpha}} = \bigotimes_{i=1}^m \ket{\alpha_i}$, $\alpha_i \in \mathbb{C}$, and a classical state takes the form
\begin{align}
    \rho_\text{cl} = \int \dd^{2m} \vb*{\alpha} \; P(\vb*{\alpha}) \dyad{\vb*{\alpha}}, \quad P(\vb*{\alpha}) \geq 0.
\end{align}
(The above decomposition can be done for any state, but it is only classical when the $P$ function is nonnegative and can thus be interpreted as a probability density function on phase space.)\\

In Ref.~\cite{Rivas2010Precision}, a upper bound on the QFI of classical states is given for any generator $H$.
We take $H = \normord{H}$, meaning that it is normally ordered -- expressed as a power series of ladder operators, all creation operators are placed on the left.
Then, for any classical $\sigma_\text{cl}$,
\begin{align} \label{eqn:cl_general_bound}
    \mc{F}(\sigma_\text{cl}, H) \leq 4 \ev{H^2 - \normord{H^2}}_{\sigma_\text{cl}}.
\end{align}
We can apply this result to Eq.~\eqref{eqn:optical_1bs}, where $\sigma_\text{cl} = \rho_\text{cl} \ox \gamma$ is classical, and $H = G$, so
\begin{align}
    G^2 - \normord{G^2} & = (a^\dagger b + b^\dagger a)^2 - \normord{(a^\dagger b + b^\dagger a)^2} \nonumber \\
        & = \left[{a^\dagger}^2 b^2 + {b^\dagger}^2 a^2 + a^\dagger b b^\dagger a + b^\dagger a a^\dagger b \right] - \left[{a^\dagger}^2 b^2 + {b^\dagger}^2 a^2 + 2 a^\dagger b^\dagger a b \right] \nonumber \\
        & = a^\dagger a (b b^\dagger - b^\dagger b) + b^\dagger b (a a^\dagger - a^\dagger a) \nonumber \\
        & = a^\dagger a + b^\dagger b .
\end{align}
Therefore,
\begin{align}
    \frac{1}{4}\mc{F}(\rho_\text{cl} \ox \gamma, G) \leq \ev{N}_{\rho_\text{cl}} + \nth,
\end{align}
which gives the following:
\begin{thm} \label{thm:classical}
    \begin{mdframed}
        For any classical state $\rho_\text{cl}$, the latent coherence is upper-bounded by
        \begin{align}
            \coh_r(\rho_\text{cl}) \leq \frac{\ev{N}_{\rho_\text{cl}} + \nth}{\nth + 1} = (1-r) \ev{N}_{\rho_\text{cl}} + r.
        \end{align}
    \end{mdframed}
\end{thm}
Comparing with Eq.~\eqref{eqn:coh_upper}, the gap between the maximal value over all states versus classical states (for fixed mean number) is $2 \nth \ev{N}/ (\nth+1) = 2r \ev{N}$.\\

It is also possible to generalise the nonclassicality witness to nonlinear generators of the form $G_k := {a^\dagger}^k b^k + a^k {b^\dagger}^k$, acting on the initial state $\rho_S \ox \gamma_B$.
By the same techniques as for $k=1$, one can easily show that
\begin{align}
     \mathcal{F}(\rho_S \ox \gamma_B, G_k) & = 4 \ev*{(N+1)^{(k)}}_{\gamma} \sum_n f(p_n, r^k p_{n-k}) n_{(k)},
\end{align}
where we introduce the rising and falling factorials, respectively $x^{(k)} = x(x+1)\dots(x+k-1)$ and $x_{(k)} = x(x-1)\dots(x-k+1)$.
From the series expansion of $(1-r)^{-1}$, one can show that $\ev*{(N+1)^{(k)}}_{\gamma} = k! (\nth+1)^k$ and $\ev*{N_{(k)}}_{\gamma} = k! \nth^k$, thus
\begin{align}
    \mathcal{F}(\rho_S \ox \gamma_B, G_k) & = 4 k! (\nth+1)^k \sum_n f(p_n, r^k p_{n-k}) n_{(k)}.
\end{align}
Aside from $k=1$, it is not clear if this quantity is a monotone under Gaussian thermal operations (as it is not defined via an optimal metrology protocol); however, it does vanish exactly when $\rho = \gamma$.

Again using Eq.~\eqref{eqn:cl_general_bound},
\begin{align}
    \ev*{G_k^2 - \normord{G_k^2}}_{\rho\ox \gamma} & = \ev{ ({a^\dagger}^k a^k) (b^k {b^\dagger}^k) + (a^k{a^\dagger}^k)({b^\dagger}^k b^k) - 2 ({a^\dagger}^k a^k)({b^\dagger}^k b^k) } \nonumber \\
        & = \ev*{N_{(k)}}_\rho k! (\nth+1)^k + \ev*{(N+1)^{(k)}}_\rho k! \nth^k - 2\ev*{N_{(k)}}_\rho k! \nth^k.
\end{align}
For any classical $\rho_\text{cl}$, one therefore has
\begin{align} \label{eqn:nonlinear_classical_bound}
    \sum_n f(p_n, r^k p_{n-k}) n_{(k)} & \leq \ev*{N_{(k)}}_\rho + \left(\frac{\nth}{\nth+1}\right)^k \left[ \ev*{(N+1)^{(k)}} - 2\ev*{N_{(k)}} \right].
\end{align}
In the low-temperature limit $r \to 0$, $\mc{F}(\rho_S \ox \gamma_B, G_k) \to 4 k! \ev*{N_{(k)}}$ for all states, so the bound Eq.~\eqref{eqn:nonlinear_classical_bound} becomes useless.
For high temperature $r \to 1$, however, we have the limiting inequality for classical states
\begin{align}
    \sum_n f(p_n, p_{n-k}) n_{(k)} \leq \ev*{(N+1)^{(k)}} - \ev*{N_{(k)}}.
\end{align}

\section{Complementarity between phase sensing and illumination} \label{app:illumination}

Quantum illumination is a detection task which makes use of quantum correlations to enhance the ability to sense a weakly reflective object~\cite{Lloyd2008Enhanced}.
If the object is not there, thermal background noise is returned to the detector; if it is there, a small amount of the signal is returned on top of the noise.
Instead of this binary discrimination problem, illumination can be also recast as a metrology protocol~\cite{Sanz2017Quantum}.
The signal is mixed with thermal light at a beam splitter with weak reflectivity $R \ll 1$, and the task is to estimate $R$.
In the limit $R \to 0$, the QFI for any single-mode phase-invariant signal vanishes.
However, by using entanglement with an external reference system (called the ``idler" $I$), nonzero sensitivity can be achieved.

We take an entangled state that is a purification of a phase-invariant state $\rho_{S}$:
\begin{align}
    \ket{\psi}_{SI} = \sum_n \sqrt{p_n} \ket{n}_{S} \ket{v_n}_{I}.
\end{align}
The output state following the beam splitter unitary $U_R$ and tracing out the thermal environment is
\begin{align}
    \zeta_{SI}(R) = \tr_B \left[U_R(\dyad{\psi}_{SI} \ox \gamma_B) U_R^\dagger \right],
\end{align}
whose QFI with respect to $R$ is denoted by $\mc{F}[\zeta_{SI}(R)]$.
Following Ref.~\cite{Sanz2017Quantum}, this QFI is
\begin{align} \label{eqn:illum_qfi}
    \mc{F}[\zeta_{SI}(R)] & = \frac{4}{\nth+1} \sum_{n,m} \frac{p_n p_m}{p_n+rp_m} \abs{\mel{m}{a}{n}}^2 \nonumber \\
        & = \frac{4}{\nth+1} \sum_n \frac{p_n p_{n-1}}{p_n + r p_{n-1}} n \nonumber \\
        & = \frac{\illum_r(\rho_S)}{\nth},
\end{align}
where we have defined the \emph{illuminance}
\begin{align} \label{eqn:illum_defn}
    \illum_r(\rho) := 4 \sum_n \frac{r p_n p_{n-1}}{p_n + rp_{n-1}} n.
\end{align}
From the identity
\begin{align}
    \frac{(x-y)^2}{x+y} + \frac{4xy}{x+y} = x+y,
\end{align}
we arrive at the following.

\begin{thm} \label{thm:tradeoff}
    \begin{mdframed}
        The latent coherence and illuminance of a phase-invariant single-mode state $\rho$ have the trade-off relation
        \begin{align}
            \coh_r(\rho) + \illum_r(\rho) = (1+r) \ev{N}_\rho + r.
        \end{align}
        This can be written as a complementarity relation between the QFI $\mc{F}_\theta$ for phase sensing with $\rho$ and the QFI $\mc{F}_R$ for quantum illumination with a purification $\ket{\psi}_{SI}$ of $\rho$:
        \begin{align} \label{eqn:qfi_complement}
            (1-r) \mc{F}_\theta + \left(\frac{r}{1-r}\right) \mc{F}_R & = (1+r) \ev{N}_\rho + r .
        \end{align}
        If we replace $\ket{\psi}_{SI}$ by a more general mixed state, then \eqref{eqn:qfi_complement} holds as an upper bound.
    \end{mdframed}
\end{thm}

\begin{figure}[h]
    \includegraphics[width=.42\textwidth]{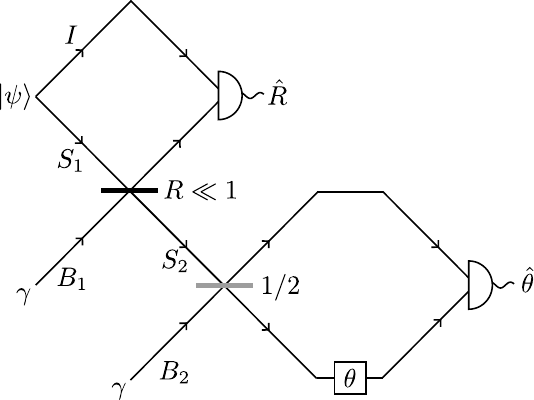}
    \caption{A single resource state $\ket{\psi}_{{S_1}I} = \sum_n \sqrt{p_n} \ket{n}_{S_1} \ket{v_n}_I$ can be used to sense both a weak reflectivity parameter $R \ll 1$ and a phase $\theta$, taking thermal light from two baths $B_1,\, B_2$. Since almost all the light from $S_1$ is passed by the first beam splitter, we identify the states of $S_1$ and $S_2$}
    \label{fig:tradeoff}
\end{figure}

This trade-off relation can be interpreted as coming from the utility of a single resource state $\ket{\psi}$ for simultaneously sensing both a phase $\theta$ and the reflectivity $R$ of a weak beam splitter -- see Fig.~\ref{fig:tradeoff}.
The part of $\ket{\psi}$ passed to the weak beam splitter is almost entirely transmitted (since $R \ll 1$), thus this light can be used for sensing the phase, in conjunction with another thermal state at a balanced beam splitter.

Note that $\illum_0 = 0$ although the corresponding QFI is nonzero: $\mc{F}_R = 4\sum_n p_{n-1}n = 4(\ev{N}_\rho +1)$ (the denominator in Eq.~\eqref{eqn:illum_qfi} vanishes).
Therefore, this complementarity relation is meaningful only at nonzero temperature.

As referenced earlier in Eq.~\eqref{eqn:coh_upper}, the maximal $\coh_r$ for a given mean number can be found by setting $\illum_r = 0$.
From Eq.~\eqref{eqn:illum_defn}, this condition is equivalent to $p_n p_{n-1} = 0 \; \forall \, n$ -- i.e., $\rho$ must contain only even or only odd photon numbers.
This is the same as $\tr(\rho a \rho a^\dagger) = 0$.
Conversely, maximal $\illum_r$ is obtained for $\coh_r=0$ -- when $\rho = \gamma$.

\section{Optical interaction speed} \label{app:speed_optical}
\begin{thm} \label{thm:optical_interaction}
    \begin{mdframed}
        The maximum interaction QFI in the linear optical setting is equivalent to the phase sensing result:
        \begin{align}
            \sup_k \max_{\substack{H_I = \vb{a}^\dagger h \vb{a}, \\ \spr(h) \leq 1}} \mc{F}(\rho_S \ox \gamma_B^{\ox k}, H_I) = (\nth + 1) \coh_r(\rho_S),
        \end{align}
        attained with $k=1$ and a beam splitter Hamiltonian $H_I = (a^\dagger b + b^\dagger a)/2$.
    \end{mdframed}
\end{thm}
To show this, we write
\begin{align}
    H_I & = \vb{a}^\dagger h \vb{a} \nonumber \\
        & = h_{00} a_0^\dagger a_0 + \left(\sum_{i,j=1}^k h_{ij} a_i^\dagger a_j \right) + \left( \sum_{i=1}^k h_{0i} a_0^\dagger a_i + h_{i0} a_i^\dagger a_0 \right),
\end{align}
and note that the first two terms do not contribute to the QFI due to phase-invariance of $\rho_S$ and full rotational invariance of $\gamma_B^{\ox k}$.
Therefore,
\begin{align}
    \mc{F}(\rho_S \ox \gamma_B^{\ox k},H_I) & = \mc{F}\left(\rho_S \ox \gamma_B^{\ox k}, \sum_{i=1}^k h_{0i} a_0^\dagger a_i + h_{i0} a_i^\dagger a_0 \right) \nonumber \\
        & = \abs{\vb{v}}^2 \mc{F}(\rho_S \ox \gamma_B^{\ox k}, a_0^\dagger b^\dagger + b^\dagger a_0),
\end{align}
where $\vb{v} = (0,h_{01},h_{02},\dots,h_{0k})^T$ and $b = \vb{v}^T \vb{a} / \abs{\vb{v}}$ is a ladder operator with the standard normalisation.
By a suitable rotation of modes, we can reduce the above to $k=1$, i.e., to $\abs{\vb{v}}^2 \mc{F}(\rho_S \ox \gamma_B, a_0^\dagger b + b^\dagger a_0)$.
The task is then to maximise $\abs{\vb{v}}$.
One characterisation of the matrix spread for Hermitian $h$ is~\cite{Johnson1985Lower}
\begin{align}
    \spr(h) = 2 \sup_{\substack{\vb{w},\vb{z} \\ \abs{\vb{w}} = \abs{\vb{z}} = 1 \\ \vb{w}^\dagger \vb{z} = 0}} \abs{\vb{w}^\dagger h \vb{z}}.
\end{align}
Taking $\vb{z} = \vb{v}^*/\abs{\vb{v}}$ and $\vb{w} = (1,0,0,\dots)^T$, it follows that
\begin{align}
    \spr(h) & \geq \frac{2}{\abs{\vb{v}}} \abs{\sum_{i,j=0}^k \delta_{i,0} h_{ij} v_j^*} \nonumber \\
        & = \frac{2}{\abs{\vb{v}}} \sum_{j=1}^k \abs{h_{0j}}^2 \nonumber \\
        & = 2 \abs{\vb{v}}.
\end{align}
Given the constraint $\spr(h) \leq 1$, we have the maximal QFI
\begin{align}
    \mc{F}(\rho_S \ox \gamma_B, H_I) & \leq \left( \frac{\spr(h)}{2} \right)^2 \mc{F}(\rho_S \ox \gamma_B, a_0^\dagger b + b^\dagger a_0) \nonumber \\
        & \leq \frac{1}{4} \mc{F}(\rho_S \ox \gamma_B, a_0^\dagger b + b^\dagger a_0),
\end{align}
which can be saturated with $h = \frac{1}{2} \left( \begin{smallmatrix} 0 & 1 \\ 1 & 0 \end{smallmatrix} \right)$.
This result is exactly as obtained in Appendix~\ref{app:max_qfi_optical}.

\section{Relation to standard optical coherence} \label{app:standard_coherence}

Here, we explore the connection between latent coherence and standard notions of quantum optical coherence~\cite{Gerry2004IntroductoryCh5}.

\subsection{First-order coherence}
In Ref.~\cite{Baker2021Heisenberg}, the coherence $\mathfrak{C}$ of a bosonic field is defined as the maximal mean photon number in any single spatial mode, restricted to some frequency band.
This can be motivated by a connection to the peak value of the power spectrum, the Fourier transform of the first-order coherence function $\ev{a^\dagger(t+\tau) a(t)}$ (using Heisenberg-picture ladder operators for a one-dimensional beam).
The restriction to a frequency band is necessary to avoid the infrared divergence which would otherwise imply divergent coherence of a thermal beam: a thermal mode at low frequency $\omega$ has $\nth = (e^{\beta \hbar \omega} -1)^{-1} \approx \kb T / \hbar \omega$.

The use of latent coherence automatically resolves the divergence issue: choosing a background at temperature $T$, for a frequency $\omega$ of a thermal beam at temperature $T'$, Eq.~\eqref{eqn:thermal_state_coherence} gives
\begin{align}
    \mathcal{C}_{e^{-\beta \hbar \omega}}(\gamma'[\omega]) & = \frac{(e^{-\beta' \hbar \omega}-e^{-\beta \hbar \omega})^2}{(e^{-\beta' \hbar \omega}+e^{-\beta \hbar \omega}) (1-e^{-\beta' \hbar \omega})}.
\end{align}
This quantity is linear in $\omega$ as $\omega \to 0$, so does not diverge for any finite temperature.
The maximum over $\omega$ can be found numerically; for example, for a room-temperature background $T=\SI{300}{K}$ and a light bulb at $T'=\SI{3000}{K}$, one obtains a maximum value of $2.72$ at an infrared wavelength $\lambda = \SI{27}{\micro \meter}$.
The mean photon number at this wavelength is about $5.1$.
In contrast, a typical optical-frequency laser has roughly $10^{12}$ photons at its peak wavelength~\cite{Wiseman2016How}, where the background thermal occupation is negligible, so then $\coh_r \sim 10^{12}$.

Note that Baker et al.~\cite[Supplementary Section 1.6]{Baker2021Heisenberg} also explore number-basis coherence generated by an interferometer (using a vacuum background).
Instead of using QFI, they argue that the relative entropy of asymmetry approaches $\frac{1}{2} \ln \mathfrak{C}$ for a single frequency mode of a laser with large photon number.

\subsection{Second-order coherence}
Second-order optical coherence describes correlations between the intensities of different modes (e.g., different spatial modes or with a time delay), considering expectation values of the form $\ev{a^\dagger b^\dagger b a}$.
For two orthogonal modes, this reduces to $\ev{N_A N_B}$; for $A=B$, it gives $\ev{N_A^2} - \ev{N_A}$.
One can form the normalised second-order coherence function~\cite{Gerry2004IntroductoryCh5}
\begin{align}
    \gt(A,B) & := \frac{\ev{a^\dagger b^\dagger b a}}{\ev{a^\dagger a} \ev{b^\dagger b}}.
\end{align}
A single mode in a coherent state has $\gt(A,A) = 1$, while thermal light has $\gt(A,A) = 2$.
A value under or over 1 is described as sub- or super-Poissonian respectively, which is justified by being equivalent to $\variance(N_A) < \ev{N_A}$ or $\variance(N_A) > \ev{N_A}$, where $\variance(N_A) = \ev{N_A^2} - \ev{N_A}^2$ is the number variance.
This property can be quantified by the $Q$-parameter~\cite{Gerry2004IntroductoryCh5},
\begin{align}
    Q(A):= \frac{\variance(N_A) - \ev{N_A}}{\ev{N_A}}.
\end{align}
The minimum value $Q = -1$ is obtained for vanishing variance, i.e., for Fock states.

We relate the input coherence to number correlations between the two branches $S$, $B$ of the interferometer.
In the standard (zero-temperature background) case, we examine the covariance between the modes $S'$, $B'$ after the beam splitter, which is easily computed as
\begin{align}
    \cov(N_{S'},N_{B'}) & := \ev{N_{S'} N_{B'}} - \ev{N_{S'}}\ev{N_{B'}} \nonumber \\
        & = Q(S) \frac{\ev{N_S}}{4}.
\end{align}
Therefore, negative and positive number correlations are associated with sub- and super-Poissonian statistics respectively.\\

These statements can all be generalised to nonzero background temperature.
Following the normalisation of the $\gt$ function, we form a normalised covariance of the modes $S'$, $B'$:
\begin{align}
    \ncov(N_{S'},N_{B'}) & := \frac{\cov(N_{S'},N_{B'})}{\ev{N_{S'}} \ev{N_{B'}}} = \gt(N_{S'}, N_{B'}) - 1 \nonumber \\
        & = \frac{\ev{N_S^2} - \ev{N_S}}{(\ev{N_S} + \nth)^2} - 1 \nonumber \\
        & = \frac{\variance(N_S) + \ev{N_S}^2 - \ev{N_S}}{(\ev{N_S} + \nth)^2} - 1.
\end{align}
In the case $\nth=0$, this recovers $\ncov(N_{S'},N_{B'}) = Q(S)/\ev{N_S}$.
We make the following observations, which generalise the concept of sub-Poissonian statistics to the thermal background setting:

\begin{itemize}
    \item There are negative number correlations between the branches, $\ncov < 0$, if and only if
        \begin{align} \label{eqn:sub-poisson_thermal}
            \variance(N_S) < (2\nth + 1)\ev{N_S} - \nth^2 .
        \end{align}
    \item The inequality \eqref{eqn:sub-poisson_thermal} holds for \emph{some} background temperature if and only if the statistics are sub-thermal, in the sense that
        \begin{align}
            \variance(N_S) < \ev{N_S} (\ev{N_S} + 1).
        \end{align}
    This can be seen be maximising the right-hand side of \eqref{eqn:sub-poisson_thermal} over $\nth$, attained at $\nth = \ev{N_S}$.
    \item For all inputs, $\ncov \geq -1$.
    The minimal value of $-1$ is achieved for states of the form $(1-p)\dyad{0} + p\dyad{1}$, $0<p\leq 1$.
    This follows from the inequality $\ev{N_S^2} \geq \ev{N_S}$, which is strict whenever numbers greater than $1$ are present.
    \item Thermal state with mean number $\bar{n}_{r'}$:
        \begin{align}
            \ncov =  \left(\frac{\nth - \bar{n}_{r'}}{\nth + \bar{n}_{r'}}\right)^2.
        \end{align}
    \item Poisson statistics with mean $\lambda$:
        \begin{align}
            \ncov = \frac{\nth (\nth - 2\lambda)}{(\nth + \lambda)^2}.
        \end{align}
    This goes to zero as $\lambda \to \infty$.
    For fixed $\nth$, the minimal value of $-1/3$ is reached with $\lambda = 2\nth$.
    \item Fock state $\ket{n}$:
        \begin{align}
            \ncov = \frac{\nth^2 - (2\nth+1)n}{(\nth+n)^2}.
        \end{align}
\end{itemize}

Negative number correlations also have metrological significance.
Due to the input state $\rho_S \ox \gamma_B$ being diagonal in the number basis, we can rewrite the QFI as
\begin{align}
    \mc{F}(U[\rho_S \ox \gamma_B]U^\dagger, N_{S'}) & = \mc{F}(U[\rho_S \ox \gamma_B]U^\dagger, [N_{S'}-N_{B'}]/2) \nonumber \\
        & \leq \variance(N_{S'} - N_{B'}) \nonumber \\
        & = \variance(N_{S'}) + \variance(N_{B'}) - 2\cov(N_{S'},N_{B'}) ,
\end{align}
since the inequality $\mc{F}(\rho,H) \leq 4 \variance(H)$ holds for all states and generators.
On the other hand, any two-mode separable state $\tau_\text{sep}$ can be shown to have the upper limit~\cite{Fadel2023Multiparameter}
\begin{align} \label{eqn:sep_qfi_bound}
    \mc{F}(\tau_\text{sep}, [N_{S'} - N_{B'}]/2) \leq \variance(N_{S'}) + \variance(N_{B'}).
\end{align}
Therefore, we see that negative number correlations are a prerequisite (but not sufficient) for having entanglement that can violate the metrological bound Eq.~\eqref{eqn:sep_qfi_bound}.

\end{document}